\documentclass[aps,pre,longbibliography,amsmath,superscriptaddress,floatfix,notitlepage]{revtex4-1}
\pdfoutput=1

\usepackage{amssymb,bm,diagbox,graphicx,multirow,tikz}

\usepackage[caption=false]{subfig}

\usepackage{newfloat}
\DeclareFloatingEnvironment[
	fileext=loa,
	listname=List of Alg.s,
	name=ALG.,
	placement=tp,
]{algorithm}
\usepackage{algpseudocode}
\algrenewcommand\algorithmicindent{1em}%
\algrenewcomment[1]{\hangindent=0.5\linewidth \hfill \parbox[t]{0.5\linewidth}{\raggedright \settowidth{\hangindent}{\(\triangleright\) }\(\triangleright\) #1}}

\usepackage{fixmath}

\usepackage{hyperref}
\hypersetup{
		linktocpage,
    colorlinks,
    citecolor=black,
    filecolor=black,
    linkcolor=black,
    urlcolor=black,
}

\renewcommand{\vec}[1]{\mathbold{#1}}

\newcommand{\E}{\mathbb{E}}

\newcommand{\mat}[1]{\mathbold{#1}}
\newcommand{\matt}[1]{\bm{\mathcal{#1}}}

\newcommand{\Pers}{\mathrm{Pers}}

\newcommand{\Pp}{\mathbb{P}}

\newcommand*{\highlight}[1]{
	\tikz[baseline=(X.base)] \node[rectangle, fill=orange!25, inner sep=0.4mm] (X) {#1};}

\graphicspath{{./img/}}

\begin{document}

\title{Relating modularity maximization and stochastic block models in multilayer networks} 

\author{A. Roxana Pamfil}
\affiliation{Mathematical Institute, University of Oxford, Oxford OX2 6GG, United Kingdom}

\author{Sam D. Howison}
\affiliation{Mathematical Institute, University of Oxford, Oxford OX2 6GG, United Kingdom}

\author{Renaud Lambiotte}
\affiliation{Mathematical Institute, University of Oxford, Oxford OX2 6GG, United Kingdom}

\author{Mason A. Porter}
\affiliation{Mathematical Institute, University of Oxford, Oxford OX2 6GG, United Kingdom}
\affiliation{Department of Mathematics, University of California, Los Angeles, Los Angeles, California 90095, USA}

\begin{abstract}

Characterizing large-scale organization in networks, including multilayer networks, is one of the most prominent topics in network science and is important for many applications. One type of mesoscale feature is community structure, in which sets of nodes are densely connected internally but sparsely connected to other dense sets of nodes. Two of the most popular approaches for community detection are to maximize an objective function called ``modularity'' and to perform statistical inference using stochastic block models. Generalizing work by Newman on monolayer networks 
(\emph{Physical Review E} {\bf 94}, 052315), we show in multilayer networks that maximizing modularity is equivalent, under certain conditions, to maximizing the posterior probability of community assignments under a suitably chosen stochastic block model. We derive versions of this equivalence for various types of multilayer structure, including temporal, multiplex, and multilevel networks. We consider cases in which the key parameters are constant, as well as ones in which they vary across layers; in the latter case, this yields a novel, layer-weighted version of the modularity function. Our results also help address a longstanding difficulty of multilayer modularity-maximization algorithms, which require the specification of two sets of tuning parameters that have been difficult to choose in practice. We show how to perform this parameter selection in a statistically-grounded way, and we demonstrate the effectiveness of our approach on both synthetic and empirical networks.

\end{abstract}

\maketitle
 

\section{Introduction}\label{sec:introduction}

Networks are useful representations of systems of interacting entities, such as cities linked by railways, computers connected on the internet, or neurons interacting through synapses \cite{newman2018book}. The simplest kind of network is a graph --- a collection of nodes (i.e., vertices) that are linked to each other, in pairwise fashion, by edges. Variants include weighted networks, in which real numbers associated to edges indicate the strength of the associated connections, and directed networks, which one can use to encode asymmetric interactions between entities. An enormous amount of research is dedicated to studying mesoscale structures in networks. Examples include community structure \cite{fortunato2016community}, core--periphery structure \cite{csermely2013,rombach2017}, and role similarity \cite{rossi2015}.

In the present paper, we focus on community detection \cite{porter2009communities,fortunato2010community,fortunato2016community,abbe2017,schaub2017,peixoto2017bayesian}. There is no consensus definition of what constitutes a community, and the precise definition typically depends on the methodology that one uses to detect communities. In qualitative terms, communities are sets of nodes that are densely connected with each other and sparsely connected with other nodes in a network. Communities can reveal meaningful structure in networks, uncovering functional pathways in metabolic networks, related pages in the World Wide Web, groups of friends in social networks, and more \cite{newman2018book,porter2009communities}. 

Many existing community-detection methods involve the optimization of some quality function, such as modularity \cite{newman2004girvan,newman2006}, Markov stability \cite{lambiotte2008,delvenne2010,lambiotte2014}, the ``map equation" in Infomap \cite{rosvall2008,dedomenico2015}, and likelihood functions or posterior probabilities from generative network models \cite{karrer2011,peixoto2013parsimonious,newman2016annotated,peixoto2018}.
Among these approaches, we distinguish two types (see also \cite{ghasemian2018}). 
The first set of methods relies on the definition of a quality function based on heuristic, information-theoretic, or other arguments. For example, one may construe a partition of a network as ``better" if its communities have a ``surprising" number of connections compared with some null model (\textit{modularity}), if they form bottlenecks to random walks (\textit{stability}), or if they give a good compression of a network from an information-theoretic perspective (\textit{Infomap}). The second set of methods encompasses generative models of networks with mesoscale structures (in particular, community structure), which one then fits to data using some form of statistical inference. These models take many forms, and it is particularly popular to use various types of stochastic block models (SBMs). A recent paper by Newman \cite{newman2016} showed that, under certain conditions, maximizing modularity is equivalent to maximizing the likelihood of a simplified SBM known as the ``planted-partition model" \cite{condon2001}. This result plays an important role in our paper, so we discuss it in detail in Sec.~\ref{sec:monolayerEquiv}.  

For many applications, it is important to move beyond ordinary graphs (i.e., ``monolayer networks'') to consider more complicated network structures, for instance by studying a collection of interrelated networks. Examples include temporal networks \cite{holme2012temporal,peixoto2015modeling,bazzi2016}, in which edges and/or nodes change in time, and multiplex networks (a generalization of edge-colored multigraphs), in which edges correspond to different types of interactions \cite{faust1994,mucha2010}. We consider these examples (and others) using the formalism of multilayer networks \cite{kivela2014,bocca2014,dedomenico2016,whatis2018}.   

Detecting communities in multilayer networks is an active area of research.
There are now several methods, with a variety of approaches \cite{mucha2010,stanley2016,debacco2017,peixoto2015,dedomenico2015,jeub2017}, and they have been employed on numerous applications. For temporal networks, multilayer community detection has been applied to the study of brain networks \cite{betzel2016}, financial correlations \cite{bazzi2016}, scientific citations \cite{hric2017}, biological contagions \cite{sarzynska2016}, and more. For multiplex networks, examples of previous studies include investigations of genes with different types of interactions \cite{cantini2015}, social support networks in Indian villages \cite{debacco2017}, and microbial interactions at different sites in the human body \cite{stanley2016}. 

Given the popularity of modularity maximization for detecting communities in monolayer networks, it is no surprise that researchers have extended it to multilayer settings. Mucha et al. \cite{mucha2010} generalized previous studies that described communities as bottlenecks to certain random-walker dynamics \cite{lambiotte2008,lambiotte2010,lambiotte2014} to derive a multilayer modularity function, which one then seeks to optimize. Multilayer modularity requires the specification of two sets of parameters, and selecting appropriate values is crucial for uncovering meaningful community structure. In the simplest case, there are exactly two parameters: the resolution $\gamma$, which influences the sizes of detected communities; and the interlayer coupling $\omega$, which influences how much detected communities change across layers  \footnote{Recently, Vaiana and Muldoon \cite{muldoon2018} examined the interplay between these two parameters. They demonstrated the existence of a multilayer {``resolution limit"}, in that modularity maximization is unable to detect community merges across layers for some values of $\gamma$ and $\omega$.}. Even in this simplified case, determining appropriate values for $\gamma$ and $\omega$ is an open problem, and interlayer coupling parameters are especially poorly understood. In the absence of other guidance, researchers typically run modularity maximization for different values of $\gamma$ and $\omega$ and then use heuristic criteria (e.g., looking for a range of values that yield qualitatively similar community structures) to select a subset of values for in-depth investigation. Some researchers have attempted more systematic approaches, and we briefly indicate two of them. Bassett et al. \cite{bassett2013} chose values of $\gamma$ and $\omega$ that maximize the difference between the modularity $Q$ and the modularity $Q^\mathrm{null}$ of a network without community structure. This procedure reveals ``cohesive regions" in the $(\gamma,\omega)$ plane in which a given network is significantly more modular than a random network with similar characteristics. Weir et al. \cite{weir2017post} proposed a different approach. Their method takes as an input a set of partitions of a network into
communities and then discards all ``non-admissible" partitions, which they construed as those that are dominated by others for any choice of $\gamma$ and $\omega$. Both of these methods require a large number of calculations to adequately sample the space of multilayer partitions (by performing modularity maximization for many combinations of $\gamma$ and $\omega$ values). 

Stochastic block models and other generative models of networks with planted mesoscale structures have also been extended to multilayer settings. However, one shortcoming of existing SBMs is that they restrict how mesoscale structure can vary across layers, and many multilayer SBMs require that communities be identical across all layers \cite{peixoto2015,peixoto2015modeling,taylor2016,zhang2016}.
Several authors have relaxed this assumption in various ways. Ghasemian et al.~\cite{ghasemian2016} described a Markov model for temporal multilayer networks in which nodes move to new communities between successive layers with probabilities given by a transition matrix. Because their goal was to study detectability thresholds \cite{moore2017} in these networks, they made several simplifications that reduce the utility of their model for inferring mesoscale structures in many empirical networks. (For example, their model produces networks in which all nodes have the same expected degree.) The degree-corrected, mixed-membership generative model of De Bacco et al. \cite{debacco2017} assumed that block structure is identical across layers, but these blocks can induce different types of structures in different layers (e.g., assortative, disassortative, core--periphery, and so on). 
Stanley et al. \cite{stanley2016} proposed a model that allows different layers to have different communities. However, their algorithm involves a clustering step that assigns layers to ``strata", such that each stratum is described by a single SBM. Consequently, their model does not include an explicit mechanism for generating different communities in different layers. Vall\`es-Catal\`a et al. \cite{valles2016multilayer} proposed a model in which different layers can have different communities, but they assumed that one observes only an aggregate monolayer network. Consequently, one cannot fit this model to multilayer network data.
Bazzi et al. \cite{bazzi2016generative} proposed a general probabilistic model of multilayer networks in which mesoscale structure can vary arbitrarily across layers. One of the potential uses of their model is in community detection, although an inference algorithm that works in this general setting remains to be developed. Nevertheless, the ideas of Bazzi et al. play an important role in the derivation of our results in Sec. \ref{sec:multilayerEquiv} and in some of the numerical experiments in Sec.~\ref{sm:sec:directed} of our supplementary materials.

The primary goal of our paper is to demonstrate that the problem of maximizing modularity in multilayer networks is related to the problem of fitting particular multilayer SBMs to network data. Our work sheds light on some of the assumptions that are implicit in multilayer modularity maximization. It also provides a general method for devising new modularity functions, including a layer-weighted modularity that is appropriate for multilayer networks with statistical properties that may be heterogeneous across the layers. A second contribution of our paper is an iterative algorithm for estimating resolution and interlayer coupling parameters when performing multilayer modularity maximization. This algorithm enables one to detect communities that can vary in arbitrary ways between layers, something that existing SBM approaches are unable to do. Throughout the paper, we work with unweighted networks, and we briefly discuss possible extensions of our results to weighted networks in Sec.~\ref{sec:conclusions}.

The rest of our paper is organized as follows. In Sec.~\ref{sec:monolayerEquiv}, we review Newman's work in \cite{newman2016} that established a connection between modularity maximization and inference in the planted-partition model for monolayer networks. In Sec.~\ref{sec:multilayerEquiv}, we show that similar results hold for certain types of multilayer networks (including temporal and multiplex networks). We then use these results to determine appropriate values for the resolution and coupling parameters in multilayer modularity. In Sec.~\ref{subsec:iterativeAlg.}, we describe an iterative algorithm for estimating these parameters. In Sec.~\ref{subsec:numericalExamples}, we demonstrate its performance on synthetic networks. We then apply our methods to a multiplex social network in Sec.~\ref{sec:lazega}, and we conclude with some ideas for future work in Sec.~\ref{sec:conclusions}. In supplementary materials, we include derivations of our equivalence result for directed and multilevel networks, additional details about our iterative parameter-estimation algorithm, and further numerical tests on synthetic networks.


\section{Newman's result for monolayer networks}\label{sec:monolayerEquiv}

Modularity maximization and inference in SBMs are two of the most widely used methods for identifying mesoscale structure in networks \cite{fortunato2016community}. In \cite{newman2016}, Newman uncovered a connection between these two methods. We briefly summarize his argument in this section before considering the more general multilayer setting from Sec.~\ref{sec:multilayerEquiv} onwards.

Modularity is an objective function that measures the ``quality" of a partition of a network into disjoint sets of nodes known as ``communities''. Larger values of modularity correspond to partitions in which more edges fall within groups than would be expected by chance, as quantified by a null model. Let $g_i$ denote the (unknown) community label of node $i$. One seeks to determine a vector $\vec{g}$ of community assignments that maximizes modularity. In practice, because this is an NP-hard problem \cite{brandes2008}, one generically finds only local optima of this objective function.

For undirected monolayer networks, the version of modularity that is used most often employs the Newman--Girvan null model \cite{newman2004girvan} (which is closely related to configuration models \cite{fosdick2018}). In this case, the modularity function $Q(\vec{g})$ compares the entries of a given network's adjacency matrix $\mat{A}$ with the expectation of the entries after uniformly randomly rewiring the edges such that the expected degrees of the nodes are preserved: 
\begin{equation}\label{eqn:modMonolayer}
	Q(\vec{g})=\frac{1}{2m}\sum_{i,j=1}^{N}{\left(A_{ij}-\gamma\frac{d_id_j}{2m}\right)\delta(g_i,g_j)}\,,
\end{equation}
where $N$ is the number of nodes in the network, $d_i$ is the degree of node $i$ (and $d_j$ is the degree of node $j$), and $m$ is the number of edges. The Kronecker delta $\delta(\cdot,\cdot)$ is equal to $1$ whenever its arguments are the same, and it is equal to $0$ when they are different. Therefore, the only terms that contribute to the sum in Eqn.~\eqref{eqn:modMonolayer} are ones in which nodes $i$ and $j$ are in the same community. The resolution parameter $\gamma$, which is equal to $1$ in the traditional formulation of modularity \cite{newman2004girvan,newman2006}, was introduced in \cite{reichardt2006} and mitigates (but does not remove) resolution issues, wherein communities below a certain size are undetectable using the original definition of modularity \cite{fortunato2007,good2010}. By tuning this resolution parameter, one can detect coarser or finer community structure in a network.

In \cite{newman2016}, Newman compared the modularity function \eqref{eqn:modMonolayer}
with likelihood functions that arise from statistical network models. Following his derivation, consider a degree-corrected SBM that is parametrized by the matrix $\mat{\Theta}=(\theta_{rs})_{1 \leq r,s \leq K}$, where $K$ is the number of communities. In SBMs, one typically needs to assume that $K$ is given (or learned through some other means, such as Bayesian inference \cite{riolo2017}), but knowing the correct value of $K$ is not necessary to establish the results of this section. The parameters $\theta_{rs}$ indicate the propensities with which nodes in community $r$ connect to nodes in community $s$. Suppose that the number of edges between two nodes, $i$ and $j$, follows a Poisson distribution with mean $d_id_j\theta_{g_ig_j}/(2m)$. The log-likelihood of the data (i.e., the adjacency matrix) $\mat{A}$ given parameters $\mat{\Theta}$ and $\vec{g}$ is then 
\begin{equation}\label{eqn:logLikelihoodMonolayer}
	\log\Pp(\mat{A}|\mat{\Theta},\vec{g}) = \frac{1}{2}\sum_{i,j=1}^N\left[ A_{ij}\log\frac{d_id_j}{2m}+A_{ij}\log\theta_{g_ig_j}  -\log(A_{ij}!)-\frac{d_id_j}{2m}\theta_{g_ig_j}\right] \,.
\end{equation}
The factor of $1/2$ is necessary to avoid double-counting of edges for undirected networks. For directed networks, one can omit this factor from the log-likelihood, because the entries $A_{ij}$ and $A_{ji}$ are independent random variables. (For multilayer networks, this difference in scaling becomes significant for reasons that we discuss in Sec.~\ref{sm:sec:directed} of our supplementary materials.)

If we know the parameters $\mat{\Theta}$, our goal is to maximize the log-likelihood \eqref{eqn:logLikelihoodMonolayer} with respect to the partition vector $\vec{g}$. We can ignore any terms in the right-hand side that do not depend on $\vec{g}$ and write
\begin{equation}\label{eqn:loglMonolayer1}
	\log\Pp(\mat{A}|\mat{\Theta},\vec{g})=\frac{1}{2}\sum_{i,j=1}^N{\left( A_{ij}\log\theta_{g_ig_j}-\frac{d_id_j}{2m}\theta_{g_ig_j} \right)} +(\mbox{const.})\,. 
\end{equation} 

Following \cite{newman2016}, we consider a type of SBM known as a ``planted-partition model" (PPM) \cite{condon2001}. In this model, the parameters $\theta_{rs}$ take one of two values: $\theta_\mathrm{in}$ for $r=s$ and $\theta_\mathrm{out}$ for $r \neq s$. When $\theta_\mathrm{in}>\theta_\mathrm{out}$, the PPM generates networks with assortative communities, for which the density of edges is higher within sets of nodes than it is between sets. We may now write
\begin{align}\label{these-identities}
	\theta_{g_ig_j} &= (\theta_\mathrm{in}-\theta_\mathrm{out})\delta(g_i,g_j)+\theta_\mathrm{out} \,, \notag \\
	\log\theta_{g_ig_j} &= (\log\theta_\mathrm{in}-\log\theta_\mathrm{out})\delta(g_i,g_j)+\log\theta_\mathrm{out}\,.
\end{align} 
Substituting Eqns.~\eqref{these-identities} into Eqn.~\eqref{eqn:loglMonolayer1} yields
\begin{equation}\label{eqn:loglMonolayer}
	\log\Pp(\mat{A}|\mat{\Theta},\vec{g}) = \frac{1}{2} \left(\log\theta_\mathrm{in}-\log\theta_\mathrm{out}\right)
	\sum_{i,j=1}^N{\left( A_{ij}-\frac{\theta_\mathrm{in}-\theta_\mathrm{out}}
		{\log\theta_\mathrm{in}-\log\theta_\mathrm{out}}\frac{d_id_j}{2m}\right)\delta(g_i,g_j)}+(\mbox{const.})\,.
\end{equation}

The connection between the modularity in Eqn.~\eqref{eqn:modMonolayer} and the log-likelihood in Eqn.~\eqref{eqn:loglMonolayer} now becomes clear. Both are functions of $\vec{g}$ that take the same form (up to additive and multiplicative constants), provided we set
\begin{equation}\label{eqn:gammaOptimal}
	\gamma=\frac{\theta_\mathrm{in}-\theta_\mathrm{out}}{\log\theta_\mathrm{in}-\log\theta_\mathrm{out}}\,.
\end{equation}
In particular, the optimal community assignments $\vec{g}$ are the same for both functions; maximizing one is the same as maximizing the other. 

The above comparison between modularity maximization and maximum-likelihood estimation (MLE) for the PPM is useful for two reasons. First, it reveals some features of modularity that otherwise are not clear, such as the fact that it detects communities that are ``statistically equivalent" in the sense that they are characterized by the same parameters $\theta_\mathrm{in}$ and $\theta_\mathrm{out}$. Second, this argument suggests that choosing the resolution $\gamma$ as in Eqn.~\eqref{eqn:gammaOptimal} is a reasonable default when detecting communities using modularity maximization. (One can estimate the unknown values $\theta_\mathrm{in}$ and $\theta_\mathrm{out}$ that appear in Eqn.~\eqref{eqn:gammaOptimal} using iterative approaches \cite{newman2016}.) This is especially relevant when one is interested in choosing a single resolution value for a network, rather than exploring multiscale community structure \cite{schaub2012}. 

A similar equivalence holds for directed and bipartite networks, and the associated resolution value is the same as the one in \eqref{eqn:gammaOptimal}. We expect that establishing an analogous result for weighted networks is considerably more difficult.


\section{Multilayer networks}\label{sec:multilayerEquiv}

In this section, we derive a connection between multilayer modularity maximization and statistical inference in certain multilayer SBMs. We start with a general discussion of multilayer modularity in Sec. \ref{subsec:multilayerModularity} and multilayer SBMs in Sec. \ref{subsec:multilayerSBMs}. We then show how these formulations are related for temporal networks in Sec. \ref{subsec:temporal} and for multiplex networks in Sec. \ref{subsec:multiplex}. In Sec. \ref{sm:sec:multilevel} of our supplementary materials, we examine a scenario in which nodes are part of a hierarchical ``multilevel" structure and thus are not restricted to represent identical entities across layers. 


\subsection{Multilayer modularity maximization}\label{subsec:multilayerModularity}

We now introduce some terminology and notation. We study ``fully-interconnected" multilayer networks \cite{kivela2014}, in which each node is present in all layers, although the edges can differ from layer to layer. (In situations where this is not the case, one can ``pad" a network with additional nodes that are not adjacent to any other nodes in their layer.) Let $T$ denote the number of layers, and let $N$ denote the number of nodes in each layer. We write $\mat{A}^t$ for the $N \times N$ adjacency matrix of the network in layer $t \in \{1,\ldots,T\}$. We assume for simplicity that the intralayer networks are undirected, and we note that a similar derivation works for directed networks. (In fact, the example in Sec.~\ref{sec:lazega} involves directed networks.) In addition to these intralayer edges, for any two distinct layers $s, t \in \{1,\ldots,T\}$, the $N \times N$ matrix $\mat{\Omega}^{st}=(\omega_{ij}^{st})_{1 \leq i,j \leq N}$ specifies the weights of interlayer connections between nodes in those layers. It is convenient to assume for simplicity that these interlayer couplings are ``diagonal" \cite{kivela2014}, so they only connect copies of the same node in different layers. 
In this case, the matrices $\mat{\Omega}^{st}$ are themselves diagonal, and we use $\omega_i^{st} \equiv \omega_{ii}^{st}$ to denote the $i^\mathrm{th}$ diagonal entry.

Detecting communities in a multilayer network amounts to assigning each node-layer pair $(i,t)$ to a community, which we label $g_i^t$. For simplicity, and unless we specify otherwise, we use the term ``node" to refer to such a node-layer tuple. (These are called ``state nodes" in some papers, such as \cite{rosvall2014,bazzi2016generative}.) On occasion, it is useful to consider all node-layer pairs that correspond to the same entity. We use the term ``physical node" to refer to this common entity, following terminology from \cite{rosvall2014} that was later adapted for multilayer networks \cite{dedomenico2015,bazzi2016generative}. 

Analogously to the monolayer case, multilayer modularity is a function that measures the quality of a multilayer partition $\vec{g}$ of the set of nodes into communities, relative to some null model. Using the standard Newman--Girvan null model within each layer \cite{newman2004girvan,newman2006}, Mucha et al. \cite{mucha2010} derived an expression for multilayer modularity that is equivalent to 
\begin{align}\label{eqn:multilayerModularity}
	Q(\vec{g})&=\sum_{t=1}^T{\sum_{i,j=1}^N{\left( A_{ij}^t-\gamma_t\frac{d_i^td_j^t}{2m_t} \right)\delta(g_i^t,g_j^t)}}+\sum_{t=1}^T\sum_{s \neq t}{\sum_{i=1}^N{\omega_i^{st}\delta(g_i^s,g_i^t)}} \notag \\
	&=Q_\mathrm{intra}(\vec{g})+Q_\mathrm{inter}(\vec{g})\,,
\end{align}
where $d_i^t$ is the degree of node $i$ in layer $t$, the quantity $m_t$ is the number of edges in layer $t$, and $g_i^t$ is the community label of node $i$ in layer $t$ (and $g_j^t$ is defined analogously). The layer-specific resolution parameters $\gamma_t$ control the importance of the null-model network relative to the observed network. This multilayer modularity is the sum of intralayer terms $Q_{\mathrm{intra}}(\vec{g})$ (which add the monolayer modularity values contributed by each layer) and interlayer terms $Q_{\mathrm{inter}}(\vec{g})$, which depend on the multilayer partition $\vec{g}$ and the coupling between the layers. One can perform community detection by finding multilayer partitions $\vec{g}$ that maximize \eqref{eqn:multilayerModularity}.  

It is helpful to introduce an alternative formulation of the modularity function in \eqref{eqn:multilayerModularity}. A convenient representation of multilayer networks is to use a \textit{supra-adjacency matrix} \cite{gomez2013diffusion,kivela2014}, which has the following block structure: 
\begin{equation*}
	\matt{A} = 
	\begin{bmatrix}
		\mat{A}^1 & \mat{\Omega}^{12} & \cdots & \mat{\Omega}^{1T} \\
		\mat{\Omega}^{21} & \mat{A}^2 & \cdots & \mat{\Omega}^{2T} \\
		\vdots & \vdots & \ddots & \vdots \\
		\mat{\Omega}^{T1} & \mat{\Omega}^{T2} & \cdots & \mat{A}^T 
	\end{bmatrix}\,.
\end{equation*}
For each layer $t$, we also define a \textit{modularity matrix} $\mat{B}^t$, with entries
\begin{equation*}
	B_{ij}^t=A_{ij}^t-\gamma_t\frac{d_i^td_j^t}{2m_t}\,. 
\end{equation*}
This results in a multilayer modularity matrix $\matt{B}$ with diagonal blocks $\mat{B}^t$ and off-diagonal blocks $\mat{\Omega}^{st}$, analogously to the supra-adjacency matrix $\matt{A}$.
Introducing global indexes for the nodes, $i'=i+(t-1)N$ (so that nodes in the first layer are labeled $1,\ldots,N$, nodes in the second layer are labeled $N+1,\ldots,2N$, and so on), the modularity function in Eqn.~\eqref{eqn:multilayerModularity} is
\begin{equation}\label{eqn:modularityBmatrix}
	Q(\vec{g})=\sum_{i',j'=1}^{NT}{B_{i'j'}\delta(g_{i'},g_{j'})}\,.
\end{equation}
This objective function is the one that is maximized in the software known as {\sc GenLouvain}~\cite{genlouvain}, a multilayer generalization of the popular Louvain method for monolayer modularity maximization \cite{blondel2008}. Having an algorithm that is able to perform community detection at this level of generality is important for some of the models that we consider in Sec.~\ref{subsec:temporal} and in the supplementary materials. Such a general algorithm also makes it possible to optimize new versions of modularity that one can derive using the methods in the present paper; we provide some suggestions in this direction in Sec.~\ref{sec:conclusions}. 


\subsection{Multilayer SBMs}\label{subsec:multilayerSBMs}

We now describe a generative model whose posterior probability recovers the modularity function in Eqn.~\eqref{eqn:multilayerModularity}. Many different types of multilayer SBMs have been studied \cite{peixoto2015,peixoto2015modeling,stanley2016,debacco2017,taylor2016,
	zhang2016,ghasemian2016,bazzi2016generative,valles2016multilayer}, and there are many more that one can propose, so we start by pointing out some requirements for such models to be suited for our purpose. 

The first key requirement is that the statistical model has a posterior probability over multilayer partitions $\vec{g}$ that one can write in closed form; otherwise, we are unable to make a meaningful comparison with multilayer modularity. This rules out the general benchmark model in \cite{bazzi2016generative}, in which one uses Monte-Carlo methods to sample from the distribution of multilayer partitions. The second requirement is that the model should allow arbitrary variations in community structure across layers. This is important, because the modularity function in Eqn.~\eqref{eqn:multilayerModularity} also allows this type of generality. This requirement rules out many existing multilayer SBMs \cite{peixoto2015,peixoto2015modeling,taylor2016,zhang2016}. Finally, we demand that any interdependencies between layers are induced solely through the multilayer partition $\vec{g}$ and the coupling between nodes in different layers, rather than through intralayer edges. This rules out, for example, the model proposed in \cite{barucca2017}, which includes an ``edge persistence" parameter that favors intralayer edges remaining active over multiple temporal layers. Note that the modularity function in Eqn.~\eqref{eqn:multilayerModularity} also encapsulates all interlayer dependencies in a separate term $Q_\mathrm{inter}(\vec{g})$, which does not involve any of the intralayer adjacency matrices $\mat{A}^t$. 

With these requirements in mind, we are ready to describe the building blocks of a multilayer SBM that recovers the modularity function in Eqn.~\eqref{eqn:multilayerModularity}. In the process, we have to place further restrictions on the generative model; this is unsurprising, because SBMs allow more general mesoscale structures than the type of communities that are detectable through modularity maximization. (In the monolayer setting from Sec. \ref{sec:monolayerEquiv}, one such restriction is that the SBM only has two edge probabilities, $\theta_\mathrm{in}$ and $\theta_\mathrm{out}$.) 

We assume that edges are placed independently in each layer, conditioned on the multilayer partition $\vec{g}$. In particular, any coupling between nodes in different layers (through interlayer edges or otherwise) has no direct influence on the placement of edges within each layer. This is in line with many existing multilayer SBMs, including those in \cite{bazzi2016generative,ghasemian2016}, which are the existing models that most closely resemble the ones that we propose in the present paper. Using a degree-corrected SBM with a set of parameters $\mat{\Theta}^t$ for each layer, the conditional log-likelihood of observing a network specified by $\matt{A}$ is
\begin{equation}\label{eqn:multilayerLikelihood}
	\log\Pp(\matt{A}|\vec{g},\mat{\Theta})=\sum_{t=1}^T{\sum_{i,j=1}^N{
			\left( A_{ij}^t\log\theta^t_{g_i^tg_j^t}-\frac{d_i^td_j^t}{2m_t}\theta^t_{g_i^tg_j^t}\right)}} + \mbox{(const.)}\,.
\end{equation} 
(For reasons that we explain in Sec. \ref{sm:sec:directed} of our supplementary materials, we omit the factor of $1/2$ in front of the double sum.)
By taking the same approach as in Sec.~\ref{sec:monolayerEquiv}, we can relate each term in Eqn.~\eqref{eqn:multilayerLikelihood} to the corresponding intralayer modularity term from Eqn.~\eqref{eqn:multilayerModularity}. Assuming that the SBM in each layer takes the form of a PPM with two parameters, $\theta_\mathrm{in}^t$ and $\theta_\mathrm{out}^t$, the log-likelihood becomes
\begin{equation}\label{eqn:multilayerLikelihoodNonunif}
	\log\Pp(\matt{A}|\vec{g},\vec{\theta}_\mathrm{in},\vec{\theta}_\mathrm{out}) 
	=\sum_{t=1}^T{\left(\log\theta_\mathrm{in}^t-\log\theta_\mathrm{out}^t\right)\sum_{i,j=1}^N{
			\left( A_{ij}^t-\gamma_t\frac{d_i^td_j^t}{2m_t}\right)\delta(g_i^t,g_j^t)}}+(\mbox{const.}) \,,
\end{equation}
where
\begin{align}\label{eqn:gammaOptimalNonunif}
	\gamma_t=\frac{\theta_\mathrm{in}^t-\theta_\mathrm{out}^t}{\log\theta_\mathrm{in}^t-\log\theta_\mathrm{out}^t}\,.
\end{align}

For now, we assume that the SBM parameters $\theta_\mathrm{in}^t$ and $\theta_\mathrm{out}^t$ are the same for all layers. (We discuss the more general setting, in which these parameters can differ across layers, in Sec. \ref{subsec:temporal}.) In this case, the log-likelihood from Eqn.~\eqref{eqn:multilayerLikelihoodNonunif} becomes
\begin{equation}\label{eqn:multilayerLikelihoodUnif}
	\log\Pp(\matt{A}|\vec{g},\theta_\mathrm{in},\theta_\mathrm{out}) 
	=\left(\log\theta_\mathrm{in}-\log\theta_\mathrm{out}\right)\sum_{t=1}^T{\sum_{i,j=1}^N{
			\left( A_{ij}^t-\gamma\frac{d_i^td_j^t}{2m_t}\right)\delta(g_i^t,g_j^t)}}+(\mbox{const.}) \,,
\end{equation}
where
\begin{align}\label{eqn:gammaOptimalUnif}
	\gamma=\frac{\theta_\mathrm{in}-\theta_\mathrm{out}}{\log\theta_\mathrm{in}-\log\theta_\mathrm{out}}\,.
\end{align}

The expression in \eqref{eqn:multilayerLikelihoodUnif} now matches the term $Q_\mathrm{intra}(\vec{g})$ from Eqn.~\eqref{eqn:multilayerModularity} (up to scaling by an expression that does not depend on $\vec{g}$). The ``optimal" resolution value from \eqref{eqn:gammaOptimalUnif} is, perhaps not surprisingly, the same as the one in Eqn.~\eqref{eqn:gammaOptimal} that was derived for the monolayer setting \cite{newman2016}. Note additionally that this derivation holds irrespective of the interdependency structure between the layers and whether a multilayer network is
temporal, multiplex, or something else. 

We have now accounted for the intralayer modularity terms in Eqn.~\eqref{eqn:multilayerModularity}. These terms, as well as the log-likelihood in Eqn.~\eqref{eqn:multilayerLikelihoodUnif}, lack any form of coupling between the community assignments $\vec{g}^t$ in different layers. Without such coupling, maximizing either expression with respect to $\vec{g}$ is no different from performing the optimization separately for each layer. We introduce layer interdependencies in our generative model through a prior distribution on $\vec{g}$. This is a prior distribution, because it encodes our belief of what community assignments are probable before observing the network data $\matt{A}$. As one might expect, the form of this prior depends markedly on the specific coupling between layers. Therefore, in our subsequent discussions, we can no longer consider the most general representations of multilayer networks. We thus focus on specific types of multilayer structure: we examine temporal networks in Sec.~\ref{subsec:temporal}, multiplex networks in Sec.~\ref{subsec:multiplex}, and multilevel networks in Sec. \ref{sm:sec:multilevel} of our supplementary materials.


\subsection{Temporal networks}\label{subsec:temporal}

In a multilayer representation of a temporal network, each layer encodes interactions that occur in some period of time. Although there are exceptions (e.g., see \cite{valdano2015}), it is common to assume that interlayer edges in these temporal networks are ``diagonal" (they only connect copies of the same node in different layers), ``layer-coupled" (their weight depends only on the layers rather than on the nodes), and ``ordinal" (they only connect successive layers) \cite{kivela2014}. We also take these interlayer edges to be directed, indicating that information flows along the arrow of time.
Let $\omega_t$ denote the weight of interlayer edges from layer $t-1$ to layer $t$, where $t \in \{2,\ldots,T\}$. The supra-adjacency matrix $\matt{A}$ then has the following structure:
\begingroup
\renewcommand*{\arraystretch}{1.5}
\begin{equation*}
	\matt{A} = 
	\begin{bmatrix}
		\mat{A}^1 & \omega_2 \mat{I} & \mbox{\textbf{0}} & \cdots & \mbox{\textbf{0}} \\
		\mbox{\textbf{0}} & \mat{A}^2 & \omega_3 \mat{I} & \cdots & \mbox{\textbf{0}} \\
		\vdots & \vdots & \vdots & \ddots & \vdots \\
		\mbox{\textbf{0}} & \mbox{\textbf{0}} & \mbox{\textbf{0}} & \cdots & \omega_T\mat{I} \\
		\mbox{\textbf{0}} & \mbox{\textbf{0}} & \mbox{\textbf{0}} & \cdots & \mat{A}^T
	\end{bmatrix}\,.
\end{equation*}
\endgroup
It is common to assume for simplicity that the parameters $\omega_2,\ldots,\omega_T$ are all equal to some value $\omega$; this is known as \textit{uniform} coupling \cite{mucha2010,bazzi2016}. In this case, the expression \eqref{eqn:multilayerModularity} for multilayer modularity becomes 
\begin{equation}\label{eqn:temporalModularity}
	Q(\vec{g})=\sum_{t=1}^T{\sum_{i,j=1}^N{\left(A_{ij}^t-\gamma\frac{d_i^td_j^t}{2m_t} \right)\delta(g_i^t,g_j^t)}} +\omega\sum_{t=2}^T{\sum_{i=1}^N{\delta(g_i^{t-1},g_i^t)}}\,. 
\end{equation}
We also assume for now that the resolution parameter $\gamma$ is the same for all layers; we will relax this assumption later in this section. After specifying values for $\gamma$ and $\omega$, modularity is a function of the community labels $\vec{g}$, where $g_i^t$ gives the community assignment of node $i$ in layer $t$. 

We relate the vectors $\vec{g}^t$ of community assignments for different layers using a prior probability $\Pp(\vec{g})$. Following previous approaches for sampling multilayer partitions \cite{bazzi2016generative,ghasemian2016}, assume that the community membership of node $i$ in layer $t$ is copied from the previous layer with probability $p$ and is sampled from some specified null distribution $\Pp_0$ with probability $1-p$. Because of the temporal structure, we sample labels in the first layer from the null distribution rather than copying them from some other layer. The probability of generating a particular multilayer partition $\vec{g}$ is then
\begin{equation}\label{eqn:tempPriorDefn}
	\Pp(\vec{g})=\Pp(\vec{g}^1)\prod_{t=2}^T{\Pp(\vec{g}^t|\vec{g}^{t-1})} 
	= \prod_{i=1}^N{\Pp_0(g_i^1)}\prod_{t=2}^T{\left\{\prod_{i=1}^N{\left[ p\delta(g_i^{t-1},g_i^{t})+(1-p)\Pp_0(g_i^t) \right]}\right\}}\,. \nonumber
\end{equation}
Taking logarithms and rearranging yields
\begin{align}\label{eqn:tempPriorExpansion}
	\log\Pp(\vec{g}) 
	&= \sum_{i=1}^N{\log\Pp_0(g_i^1)}+\sum_{t=2}^T\sum_{i=1}^N{\log\left[1+\frac{p\delta(g_i^{t-1},g_i^t)}{(1-p)\Pp_0(g_i^t)}\right]}  +\sum_{t=2}^T\sum_{i=1}^N{\log\left[(1-p)\Pp_0(g_i^t)\right]} \nonumber \\
	&= \sum_{i=1}^N{\log\Pp_0(g_i^1)} +\sum_{t=2}^T\sum_{i=1}^N{\log\left[1+\frac{p}{(1-p)\Pp_0(g_i^t)}\right]\delta(g_i^{t-1},g_i^t)} \\ &\qquad+\sum_{t=2}^T\sum_{i=1}^N{\log\left[(1-p)\Pp_0(g_i^t)\right]}\,. \nonumber
\end{align}
The second equality follows from noting that any logarithmic term in the second sum is $0$ whenever $\delta(g_i^{t-1},g_i^t)=0$. The ability to move the Kronecker delta outside 
the logarithmic expression is the key technical trick that allows us to relate this prior probability to the interlayer modularity terms from Eqn.~\eqref{eqn:multilayerModularity}. This is also the crucial step that makes certain cases, such as multiplex networks (see Sec.~\ref{subsec:multiplex}), more difficult to analyze.

Comparing the expression in Eqn.~\eqref{eqn:tempPriorExpansion} with the interlayer modularity terms in Eqn.~\eqref{eqn:temporalModularity}, we start to see a resemblance: both involve weighted sums over nodes $i$ and layers $t$ of Kronecker deltas of the form $\delta(g_i^{t-1},g_i^t)$. To obtain an exact match, we have to make further assumptions. Suppose that the null distribution $\Pp_0$ is uniform over community assignments. Specifically, if there are $K$ communities (possibly including empty ones), let $\Pp_0(g_i^t)=1/K$ for any node $i$ and layer $t$, so that each node is equally likely to be assigned to each of the communities. The first and third terms in Eqn.~\eqref{eqn:tempPriorExpansion} become constants, and the only term that depends on $\vec{g}$ is the second one. From now on, we also write $\log\Pp(\vec{g}|p,K)$ for the log-prior to emphasize its dependence on the parameters $p$ and $K$.

We can now use Bayes' rule to combine intralayer and interlayer terms that arise in our proposed multilayer SBM into a single posterior probability:
\begin{align}\label{eqn:tempPosterior}
	\log\Pp(\vec{g}|\matt{A},\theta_\mathrm{in},\theta_\mathrm{out},p,K) 
	&= \log\left(\frac{\Pp(\matt{A},\vec{g}|\theta_\mathrm{in},\theta_\mathrm{out},p,K)}{\Pp(\matt{A}|\theta_\mathrm{in},\theta_\mathrm{out},p,K)}\right)  \notag \\
	&= \log\left(\frac{\Pp(\matt{A}|\vec{g},\theta_\mathrm{in},\theta_\mathrm{out})\Pp(\vec{g}|p,K)}{\Pp(\matt{A}|\theta_\mathrm{in},\theta_\mathrm{out},p,K)}\right) \nonumber \\
	&= \log\Pp(\matt{A}|\vec{g},\theta_\mathrm{in},\theta_\mathrm{out})+\log\Pp(\vec{g}|p,K) -\log\Pp(\matt{A}|\theta_\mathrm{in},\theta_\mathrm{out},p,K) \,.
\end{align}
The first two terms of \eqref{eqn:tempPosterior} correspond, respectively, to the log-likelihood from Eqn.~\eqref{eqn:multilayerLikelihoodUnif} and the log-prior from Eqn.~\eqref{eqn:tempPriorExpansion}. The third term is difficult to compute in practice, but it does not depend on $\vec{g}$, so we can treat it as a constant for our purposes. We can also ignore multiplicative constants to obtain
\begin{align}\label{eqn:tempPosterior2}
	\log\Pp(\vec{g}|\matt{A},\theta_\mathrm{in},\theta_\mathrm{out},p,K)
	&\propto \sum_{t=1}^T{\sum_{i,j=1}^N{
			\left( A_{ij}^t-\gamma\frac{d_i^td_j^t}{2m_t}\right)\delta(g_i^t,g_j^t)}} \notag \\
	&\qquad +\sum_{t=2}^T\sum_{i=1}^N{\frac{\log\left(1+\frac{p}{1-p}K\right)}{\log\theta_\mathrm{in}-\log\theta_\mathrm{out}}\delta(g_i^{t-1},g_i^t)}+\mbox{(const.)}\,. 
\end{align}
By maximizing the right-hand side of \eqref{eqn:tempPosterior2} with respect to $\vec{g}$, we find the mode of the posterior distribution over multilayer partitions, given the multilayer network $\matt{A}$ and model parameter values $\theta_\mathrm{in}$, $\theta_\mathrm{out}$, $p$, and $K$. We can now ensure that maximizing multilayer modularity \eqref{eqn:temporalModularity} is equivalent to maximizing the posterior probability \eqref{eqn:tempPosterior2} by choosing the following values for the parameters $\gamma$ and $\omega$:
\begin{align}
	\gamma &= \frac{\theta_\mathrm{in}-\theta_\mathrm{out}}{\log\theta_\mathrm{in}-\log\theta_\mathrm{out}}\,, \label{eqn:gammaOptimalTemporal} \\
	\omega &= \frac{1}{\log\theta_\mathrm{in}-\log\theta_\mathrm{out}}\log\left(1+\frac{p}{1-p}K\right) \label{eqn:omegaOptimalTemporal} \,.
\end{align}

It is useful to point out some limiting cases for the interlayer coupling parameter $\omega$. When $p=0$, which entails that there is no copying of community assignments from one layer to the next, Eqn.~\eqref{eqn:omegaOptimalTemporal} gives $\omega=0$. This implies that community detection proceeds independently for each layer. When $p \rightarrow 1$, we obtain $\omega \rightarrow \infty$, which ensures that maximizing modularity recovers the same communities in each layer, as it should. 
Perhaps less obvious is the dependence on $\theta_\mathrm{in}$ and $\theta_\mathrm{out}$. Consider the ratio $\epsilon=\theta_\mathrm{out}/\theta_\mathrm{in}$, which determines the strength of the planted community structure. According to Eqn.~\eqref{eqn:omegaOptimalTemporal}, a weaker community structure in a network (i.e., a larger value of $\epsilon$) should entail a  larger value of $\omega$. One explanation is that by overemphasizing persistence, we take advantage of correlations between layers to increase the ``signal" of underlying community structure, while simultaneously dialing down layer-specific ``noise". Finally, $\omega$ increases (weakly) with the number of communities $K$. For fixed $\theta_\mathrm{in}$ and $\theta_\mathrm{out}$, it is typically more difficult to detect a larger number of communities than a smaller one (i.e., the detectability threshold moves up \cite{ghasemian2016}). Increasing the amount of coupling improves community-detection results by taking advantage of correlations between layers. 

Thus far, we have assumed that all layers have the same resolution parameter $\gamma$ and are connected by interlayer edges with the same weight $\omega$. This amounts to assuming that the layers are ``statistically similar", in that they can be described by a common set of parameters $\theta_\mathrm{in}$, $\theta_\mathrm{out}$, $p$, and $K$. For many applications, this assumption is unrealistic, so it is important to investigate what happens when these parameters are layer-dependent.

Proceeding in analogy to the layer-independent case, the log-likelihood of observing a network $\matt{A}$ conditioned on a multilayer partition $\vec{g}$ is
\begin{align}\label{eqn:temporalLikelihoodNonUnif}
	\log\Pp(\matt{A}|\vec{g},\vec{\theta}_\mathrm{in},\vec{\theta}_\mathrm{out})   &=\sum_{t=1}^T\Biggl[\left(\log\theta_\mathrm{in}^t-\log\theta_\mathrm{out}^t\right)
	\sum_{i,j=1}^N{\left( A_{ij}^t-\frac{\theta_\mathrm{in}^t-\theta_\mathrm{out}^t}{\log\theta_\mathrm{in}^t-\log\theta_\mathrm{out}^t}\frac{d_i^td_j^t}{2m_t}\right)\delta(g_i^t,g_j^t)\Biggr]} \nonumber \\
	&\qquad+ \mbox{(const.)}\,. 
\end{align}
To impose a prior on the multilayer partition $\vec{g}$, let $p_t$ be the probability that a node in layer $t$ copies its community label from layer $t-1$, where $t \in \{2,\ldots,T\}$. Recall that $\Pp_0$ is the null distribution for community labels when nodes do not copy their community assignments from the previous layer. We now make this distribution layer-dependent, so $\Pp_0^t$ is the null distribution for layer $t$, to allow layers with non-uniform characteristics. As before, suppose that each null distribution $\Pp_0^t$ is uniform, which entails that $\Pp_0^t(g_i^t)=1/K_t$, where $K_t$ is the number of communities (possibly including empty ones) in layer $t$. Ignoring any terms that no longer depend on $\vec{g}$, the log-prior is then
\begin{align}\label{eqn:tempPriorNonUnif}
	\log\Pp(\vec{g}) 
	&= \sum_{i=1}^N{\log\left[\Pp_0^t(g_i^1)\right]}
	+\sum_{t=2}^T\sum_{i=1}^N{\log\left[ p_t\delta(g_i^{t-1},g_i^t)+(1-p_t)\Pp_0^t(g_i^t) \right]} \notag \\
	&= \sum_{t=2}^T\sum_{i=1}^N{\log\left(1+\frac{p_t}{1-p_t}K_t\right)\delta(g_i^{t-1},g_i^t)}+ \mbox{(const.)}\,.
\end{align}

Combining Eqns.~\eqref{eqn:temporalLikelihoodNonUnif} and \eqref{eqn:tempPriorNonUnif} and multiplying by a constant for notational convenience, we see that $\vec{g}$ has
the posterior probability distribution
\begin{align}\label{eqn:tempPosteriorNonUnif}
	&\log\Pp(\vec{g}|\matt{A},\vec{\theta}_\mathrm{in},\vec{\theta}_\mathrm{out},\vec{p},\vec{K})  \\
	&\qquad\propto \sum_{t=1}^T\Biggl[\frac{\log\theta_\mathrm{in}^t-\log\theta_\mathrm{out}^t}{\langle \log\theta_\mathrm{in}^t-\log\theta_\mathrm{out}^t \rangle_t} 
	\sum_{i,j=1}^N{\left( A_{ij}^t-\frac{\theta_\mathrm{in}^t-\theta_\mathrm{out}^t}{\log\theta_\mathrm{in}^t-\log\theta_\mathrm{out}^t}\frac{d_i^td_j^t}{2m_t}\right)\delta(g_i^t,g_j^t)}\Biggr] \notag \\
	&\qquad\qquad+\sum_{t=2}^T\sum_{i=1}^N{\frac{\log\left(1+\frac{p_t}{1-p_t}K_t\right)}{\langle \log\theta_\mathrm{in}^t-\log\theta_\mathrm{out}^t\rangle_t}\delta(g_i^{t-1},g_i^t)}  + \mbox{(const.)}\,, \notag
\end{align}
where $\langle \cdot \rangle_t$ denotes a mean across all layers $t \in \{1,\ldots,T\}$.

Guided by \eqref{eqn:tempPosteriorNonUnif}, we define temporal modularity for the case with non-uniform parameters with the following formula:
\begin{equation}\label{eqn:temporalModularityNonUnif}
	Q(\vec{g})=\sum_{t=1}^T{\beta_t\sum_{i,j=1}^N{\left(A_{ij}^t-\gamma_t\frac{d_i^td_j^t}{2m_t} \right)\delta(g_i^t,g_j^t)}} +\sum_{t=2}^T{\omega_t\sum_{i=1}^N{\delta(g_i^{t-1},g_i^t)}}\,.
\end{equation}
In contrast to previous definitions of multilayer modularity \cite{mucha2010}, our definition includes a new set of layer weights $\beta_t$, in addition to the resolution parameters $\gamma_t$ and interlayer couplings $\omega_t$. Comparing \eqref{eqn:temporalModularityNonUnif} with the log-probability in Eqn.~\eqref{eqn:tempPosteriorNonUnif}, we see that these parameters should take the following values:
\begin{align}
	\gamma_t &= \frac{\theta_\mathrm{in}^t-\theta_\mathrm{out}^t}{\log\theta_\mathrm{in}^t-\log\theta_\mathrm{out}^t} \,, \label{eqn:gammaOptimalNonUnif} \\
	\omega_t &= \frac{1}{\langle \log\theta_\mathrm{in}^t-\log\theta_\mathrm{out}^t\rangle_t}\log\left(1+\frac{p_t}{1-p_t}K_t\right) \,, \label{eqn:omegaOptimalNonUnif} \\
	\beta_t &= \frac{\log\theta_\mathrm{in}^t-\log\theta_\mathrm{out}^t}{\langle \log\theta_\mathrm{in}^t-\log\theta_\mathrm{out}^t \rangle_t} \label{eqn:betaOptimalNonUnif} \,.
\end{align} 
The expressions in \eqref{eqn:gammaOptimalNonUnif}--\eqref{eqn:betaOptimalNonUnif} are consistent with the ones that we obtained for the case of uniform coupling. If the parameters $\theta_\mathrm{in}$, $\theta_\mathrm{out}$, $p_t$, and $K_t$ are layer-independent, the first two expressions reduce to Eqns.~\eqref{eqn:gammaOptimalTemporal} and \eqref{eqn:omegaOptimalTemporal}. Additionally, Eqn.~\eqref{eqn:betaOptimalNonUnif} gives $\beta_t=1$, which is why these weights do not appear explicitly when one considers uniform coupling.

The role of the parameters $\beta_t$ is to give different weightings to intralayer modularity terms that correspond to different layers. To our knowledge, this is the first definition of multilayer modularity that incorporates such layer weightings explicitly \footnote{One can write the most general version of multilayer modularity in \cite{mucha2010} in the form of Eqn.~\eqref{eqn:temporalModularityNonUnif} by rescaling the adjacency matrices $\mat{A}^t$ to incorporate the weights $\beta_t$. However, this scenario is not one that the authors of \cite{mucha2010} (or other researchers) have considered explicitly.}.
Their values depend on the strength of the community structure in their respective layers. Letting $\epsilon_t=\theta_\mathrm{out}^t/\theta_\mathrm{in}^t$, we see that $\beta_t \propto \log(1/\epsilon_t)$. This implies that layers with strong community structure (i.e., $\epsilon_t$ close to $0$) get a larger weight when maximizing multilayer modularity than layers in which the community structure is weak (i.e., $\epsilon_t$ close to $1$).
It is also worth noting what happens when some layers have disassortative structure (i.e., when $\epsilon_t>1$). The formula \eqref{eqn:betaOptimalNonUnif} suggests taking $\beta_t<0$, pushing the method towards finding multilayer partitions that \textit{minimize} monolayer modularity for these layers. Indeed, it is known that minimizing modularity for monolayer networks identifies partitions with many edges between groups and few edges within groups \cite{newman2006,newman2016}. The multilayer setting is especially interesting, because it allows (at least in principle) a combination of assortative and disassortative layers \cite{debacco2017}.

Although the proposed multilayer modularity from Eqn.~\eqref{eqn:omegaOptimalNonUnif} takes a non-standard form because of the weights $\beta_t$, methods such as {\sc GenLouvain} \cite{genlouvain}, which are written as maximization problems of the form given in Eqn.~\eqref{eqn:modularityBmatrix}, will continue to work. We also note that one can explicitly consider situations in which the interlayer coupling parameters are uniform and the resolution parameters are layer-dependent, or the other way around. 

A key facet of the results of this section is that they give a principled method for choosing heterogeneous interlayer coupling weights $\omega_t$. Although the original derivation of multilayer modularity in \cite{mucha2010} is general enough to include scenarios with non-uniform parameters, most computations that researchers have done in practice have used uniform coupling and resolution values \cite{mucha2010,bassett2013,weir2017post}. For most applications, this is a simplifying assumption rather than a hypothesis that these parameters are in fact the same for all layers. Our work suggests ways to remove this assumption and perform computations when parameters are layer-dependent, which we expect to be relevant for almost all applications.

The derivation from this section holds whenever the graph of dependencies between layers is a tree. In that case, community labels propagate from the root of this tree to the leaves, yielding a similar prior distribution $\Pp(\vec{g})$ to the one from Eqn.~\eqref{eqn:tempPriorDefn}. One example of potential interest for applications is the star graph, in which one central layer influences community structure in all remaining layers. 


\subsection{Multiplex networks}\label{subsec:multiplex}

In a multiplex network, nodes are adjacent to each other via multiple types of edges; each edge type corresponds to a layer in the resulting multilayer network \cite{kivela2014,faust1994}. In this section, we establish a relationship between modularity maximization and certain multilayer SBMs for multiplex networks.

As in Sec.~\ref{subsec:temporal}, we continue to assume that our multilayer networks are fully interconnected, diagonal, and layer-coupled. To construct a multiplex SBM, we follow \cite{bazzi2016generative} and let $\mat{P}$ denote the matrix of interlayer dependencies, where $p_{st}$ (with $s \neq t$) is the probability that a node in layer $t$ copies its community label from layer $s$. (We assume that $p_{ss}=0$.) The matrix $\mat{P}$ may or may not be symmetric. We say that the case in which all probabilities $p_{st}$ are equal to the same value $p$ has ``uniform coupling''. 

Recall from Sec. \ref{subsec:multilayerSBMs} that the intralayer modularity $Q_\mathrm{intra}(\vec{g})$ takes the same functional form as the conditional log-likelihood $\Pp(\matt{A}|\vec{g},\vec{\theta}_\mathrm{in},\vec{\theta}_\mathrm{out})$, regardless of the dependency structure between layers. We therefore focus our attention on devising a prior $\Pp(\vec{g})$ that recovers the interlayer modularity $Q_\mathrm{inter}(\vec{g})$ for multiplex networks.

In a multiplex network, there is no natural ordering of the layers (in contrast to temporal and multilevel networks). Consider the vector $\vec{g}_i=(g_i^1,\ldots,g_i^T)$ of community labels for some node $i$. From the definition of conditional probability, it follows that
\begin{equation*}
	\Pp(\vec{g}_i)=\Pp(g_i^1)\Pp(g_i^2|g_i^1)\times\cdots\times\Pp(g_i^T|g_i^1,\ldots,g_i^{T-1})\,.
\end{equation*}
However, we can also write
\begin{equation*}
	\Pp(\vec{g}_i)=\Pp(g_i^T)\Pp(g_i^{T-1}|g_i^T)\times\cdots\times\Pp(g_i^1|g_i^T,\ldots,g_i^2)\,,
\end{equation*}
and we can write a similar expression for any other permutation of the $T$ layers. Of course, all of these products have to evaluate to the same quantity [namely, $\Pp(\vec{g}_i)$]. It is difficult to define a probabilistic model for $\vec{g}$ that is internally consistent in this way. In particular, a model similar to the one in \cite{bazzi2016} that is closed-form does not have these properties.

We take a different approach. Considering each node $i$ independently, we first sample a permutation $\sigma$ of the layers from the symmetric group $S_T$ with probability $\Pp(\sigma)$, and we then update community labels in the order indicated by this permutation: first, we sample a community label for $i$ in what is now layer $1$; we then sample a label in layer $2$, conditioned on the label from layer $1$; and so on \footnote{Alternatively, one can consider sampling over all possible spanning trees with $T$ nodes and assume that the update propagates from the root of each tree to the leaves.}. The permutation $\sigma$ that indicates the update order can be different for different nodes. The resulting prior is
\begin{equation}\label{eqn:multiplexPriorInitial}
	\Pp(\vec{g})=\prod_{i=1}^N\sum_{\sigma \in S_T}{\Pp(\sigma)\Pp\left(g_i^{\sigma^{-1}(1)}\right)\prod_{t=2}^T{\Pp\left(g_i^{\sigma^{-1}(t)}|g_i^{\sigma^{-1}(t-1)}\right)}}\,.
\end{equation}
Note that we assume that the update process is memoryless, so the community label in layer $t$ (after applying the permutation $\sigma$) depends only on the community label in layer $t-1$. This is the same assumption that we (and others \cite{ghasemian2016,bazzi2016generative}) have made for temporal networks.

For simplicity, we write $\varsigma_t$ for $\sigma^{-1}(t)$. As in Sec.~\ref{subsec:temporal}, we use a copying and resampling process to generate community labels. The conditional probabilities in Eqn.~\eqref{eqn:multiplexPriorInitial} are then 
\begin{equation*}
	\Pp(g_i^{\varsigma_t}|g_i^{\varsigma_{t-1}})=p_{\varsigma_{t-1}\varsigma_t}\delta(g_i^{\varsigma_{t-1}},g_i^{\varsigma_t})+(1-p_{\varsigma_{t-1}\varsigma_t})\Pp_0^{\varsigma_t}(g_i^{\varsigma_t})\,.
\end{equation*}
For the first layer,
\begin{equation*}
	\Pp(g_i^{\varsigma_1})=\Pp_0^{\varsigma_1}(g_i^{\varsigma_1})\,.
\end{equation*}
With a uniform prior $\Pp_0^{\varsigma_t}(g_i^{\varsigma_t})=1/K_{\varsigma_t}$, we obtain
\begin{equation*}
	\Pp(\vec{g}) = \prod_{i=1}^N\sum_{\sigma \in S_T}\Biggl\{\Pp(\sigma)\frac{\prod_{t=2}^T{(1-p_{\varsigma_{t-1}\varsigma_t})}}{\prod_{t=1}^T{K_{\varsigma_t}}}
	\prod_{t=2}^T{\left[ 1+\frac{p_{\varsigma_{t-1}\varsigma_t}}{1-p_{\varsigma_{t-1}\varsigma_t}}K_{\varsigma_t}\delta(g_i^{\varsigma_{t-1}},g_i^{\varsigma_t}) \right]}\Biggr\}
\end{equation*}
Taking logarithms and treating any terms that do not depend on $\vec{g}$ as constant yields
\begin{equation}\label{eqn:multiplexPrior}
	\log\Pp(\vec{g})=\sum_{i=1}^N\log\Biggl\{\sum_{\sigma \in S_T}\Pp(\sigma)\prod_{t=2}^T(1-p_{\varsigma_{t-1}\varsigma_t}) 
	\left[ 1+\frac{p_{\varsigma_{t-1}\varsigma_t}}{1-p_{\varsigma_{t-1}\varsigma_t}}K_{\varsigma_t}\delta(g_i^{\varsigma_{t-1}},g_i^{\varsigma_t}) \right]\Biggr\}+\mbox{(const.)} \,.
\end{equation}
As we mentioned in Sec.~\ref{subsec:temporal}, being able to take the Kronecker delta outside 
the logarithmic expression is the key step that allows us to relate this prior probability to the corresponding interlayer modularity terms. However, we cannot do this easily in the present case because of the sum over permutations $\sigma \in S_T$. Instead, we bound the expression from Eqn.~\eqref{eqn:multiplexPrior} using Jensen's inequality. Because the logarithm is a concave function, 
\begin{equation}\label{jensen}
	\log\left(\sum_i{x_i}\right) \geq \sum_i{q_i\log\left(\frac{x_i}{q_i}\right)} 
\end{equation}
for any $q_i \geq 0$ with $\sum_i{q_i}=1$, and equality holds for $q_i=x_i/\sum_i{x_i}$.
Applying the inequality \eqref{jensen} to Eqn.~\eqref{eqn:multiplexPrior} and ignoring additive terms that do not depend on $\vec{g}$ yields
\begin{equation}\label{eqn:multiplexPriorBound}
	\log\Pp(\vec{g}) \geq \sum_{i=1}^N\sum_{\sigma \in S_T}q_i^\sigma\sum_{t=2}^T\log\left(1+\frac{p_{\varsigma_{t-1}\varsigma_t}}{1-p_{\varsigma_{t-1}\varsigma_t}}K_{\varsigma_t}\right)\delta(g_i^{\varsigma_{t-1}},g_i^{\varsigma_t}) + \mbox{(const.)} \,.
\end{equation}
We want to rearrange the right-hand side of \eqref{eqn:multiplexPriorBound} into an expression that resembles the interlayer modularity terms in Eqn.~\eqref{eqn:multilayerModularity}. To do this, we fix $s$ and $t$ and consider all of the terms in \eqref{eqn:multiplexPriorBound} that involve $\delta(g_i^s,g_i^t)$. There are $(T-1)!$ permutations $\sigma$ with $\varsigma_{t'-1}=s$ and $\varsigma_{t'}=t$ for some $t' \in \{2,\ldots,T\}$. For convenience, let $S_T^{(s,t)}$ denote the subset of $S_T$ that consists of permutations that map layers $s$ and $t$ to two successive layers. The lower bound of the log-prior then becomes
\begin{equation}\label{eqn:multiplexPriorBound2}
	\log\Pp(\vec{g}) \geq \sum_{i=1}^N\sum_{s \neq t}\sum_{\sigma \in S_T^{(s,t)}}{q_i^\sigma}\log\left(1+\frac{p_{st}}{1-p_{st}}K_t\right)\delta(g_i^s,g_i^t) + \mbox{(const.)} \,.
\end{equation}
The fact that we have a lower bound for this log-prior, rather than an equality as in Sec.~\ref{subsec:temporal}, may seem problematic. However, recall that we obtain equality if the quantities $q_i^\sigma$ take the specific values
\begin{equation}\label{eqn:qiSigma}
	q_i^\sigma=\frac{ \Pp(\sigma)\displaystyle\prod_{t=2}^T{(1-p_{\varsigma_{t-1}\varsigma_t})\left[ 1+\frac{p_{\varsigma_{t-1}\varsigma_t}}{1-p_{\varsigma_{t-1}\varsigma_t}}K_{\varsigma_t}\delta(g_i^{\varsigma_{t-1}},g_i^{\varsigma_t}) \right]}}{Z_i}\,,
\end{equation}
where for each node $i \in \{1,\ldots,N\}$, $Z_i$ is a proportionality constant that ensures that $\sum_{\sigma \in S_T}{q_i^\sigma}=1$. One can interpret each $q_i^\sigma$ as the posterior probability $\Pp(\sigma|\vec{g}_i)$ that we sampled the community labels for node $i$ 
in the layer order indicated by $\sigma$. This posterior probability is proportional to the product of the prior probability $\Pp(\sigma)$ and the likelihood $\Pp(\vec{g}_i|\sigma)$.

By comparing the right-hand side of \eqref{eqn:multiplexPriorBound2} with the intralayer modularity terms from \eqref{eqn:multilayerModularity}, we arrive at the following optimal values for the interlayer coupling parameters:
\begin{equation}\label{eqn:optimalOmegaMultiplexGeneral}
	\omega_{st}=\frac{1}{\langle \log\theta_\mathrm{in}^t-\log\theta_\mathrm{out}^t\rangle_t}\log\left(1+\frac{p_{st}}{1-p_{st}}K_t\right)\sum_{\sigma \in S_T^{(s,t)}}{q_i^\sigma}\,.
\end{equation}

Given a multilayer partition $\vec{g}$ and values for the other parameters of the generative model, one can (in principle) calculate values for the weights $q_i^\sigma$ using Eqn.~\eqref{eqn:qiSigma}. Substituting these values into Eqn.~\eqref{eqn:optimalOmegaMultiplexGeneral} then gives estimates for the coupling weights $\omega_{st}$. In practice, the sum over permutations $\sigma \in S_T^{(s,t)}$ from Eqn.~\eqref{eqn:optimalOmegaMultiplexGeneral} has $(T-1)!$ terms. Although there are many empirical data sets with a sufficiently small number of layers (say, $T \lessapprox 6$) for which this calculation is feasible, this brute-force approach will not work in general.

We can make some simplifications in the case of uniform coupling, for which $p_{st}=p$ for $s,t \in \{1,\ldots,T\}$. 
Furthermore, we assume that the number of communities is the same for each layer ($K_t=K$) and that we sample each permutation $\sigma$ 
with equal probability (i.e., $\Pp(\sigma)=1/T!$). 
In this case, the layers of our multiplex network are statistically equivalent, so on average any permutation of the layers should produce the same results. This implies that the quantities $q_i^\sigma=\Pp(\sigma|\vec{g}_i)$ ($\sigma \in S_T$) are equal in expectation, which yields
\begin{equation}\label{eqn:multiplexqApprox}
	\sum_{\sigma \in S_T^{(s,t)}}{q_i^\sigma} \approx \frac{1}{T}\sum_{\sigma \in S_T}{q_i^\sigma}=\frac{1}{T}\,.
\end{equation}
It is possible to make this argument mathematically rigorous in the limit $T \rightarrow \infty$ by approximating the quantities $q_i^\sigma$ (for $\sigma \in S_T^{(s,t)}$) with quantities $q_i^{\sigma'}$ (for $\sigma' \in S_T$) of similar magnitude. In Table~\ref{table:multiplexApprox}, we provide some numerical evidence that our approximation is reasonable even for small $T$. Indeed, we observe that the sample mean of the expression $\left(\sum_{\sigma \in S^{(s,t)}_T}{q_i^\sigma}\right)-1/T$ is close to $0$ for different values of the copying probability $p$ and the number of layers $T$. (For these examples, we use $K=5$.) From a practical standpoint, we expect that any errors in the approximation \eqref{eqn:multiplexqApprox} will have a small impact on the results of modularity maximization compared to other sources of error (e.g., the algorithm identifying different local optima of the modularity function across different runs).

\begin{table}[ht!]
	\centering
	\label{table:multiplexApprox}
	\caption{Sample mean and standard deviation (in parentheses) of $\sum_{\sigma \in S^{(s,t)}_T}{q_i^\sigma}-1/T$ across $50$ trials}
	\begin{tabular}{c|rrrrrrrr}
		\hline\hline
		& \multicolumn{1}{c}{$T=3$} & \multicolumn{1}{c}{$T=4$} & \multicolumn{1}{c}{$T=5$} & \multicolumn{1}{c}{$T=6$} & \multicolumn{1}{c}{$T=7$} & \multicolumn{1}{c}{$T=8$} & \multicolumn{1}{c}{$T=9$} & \multicolumn{1}{c}{$T=10$} \\ \hline
		\multirow{ 2}{*}{$p=0.5$} & $-0.013\phantom{)}$ & $0.027\phantom{)}$ & $0.009\phantom{)}$ & $0.008\phantom{)}$ & $-0.013\phantom{)}$ & $0.005\phantom{)}$ & $-0.004\phantom{)}$ & $-0.020\phantom{)}$ \\
		& $(0.052)$ & $(0.098)$ & $(0.090)$ & $(0.090)$ & $(0.076)$ & $(0.078)$ & $(0.069)$ & $(0.047)$ \\ \hline
		\multirow{ 2}{*}{$p=0.7$} & $0.000\phantom{)}$ & $-0.001\phantom{)}$ & $0.016\phantom{)}$ & $0.003\phantom{)}$ & $-0.008\phantom{)}$ & $-0.003\phantom{)}$ & $0.025\phantom{)}$ & $0.010\phantom{)}$ \\
		& $(0.068)$ & $(0.073)$ & $(0.082)$ & $(0.096)$ & $(0.071)$ & $(0.086)$ & $(0.100)$ & $(0.082)$ \\ \hline    
		\multirow{ 2}{*}{$p=0.9$} & $-0.003\phantom{)}$ & $0.006\phantom{)}$ & $-0.004\phantom{)}$ & $0.007\phantom{)}$ & $0.004\phantom{)}$ & $0.010\phantom{)}$ & $0.008\phantom{)}$ & $-0.007\phantom{)}$ \\  
		& $(0.016)$ & $(0.048)$ & $(0.064)$ & $(0.075)$ & $(0.051)$ & $(0.080)$ & $(0.059)$ & $(0.052)$ \\ \hline\hline
	\end{tabular}
\end{table}

The approximation \eqref{eqn:multiplexqApprox} gives 
\begin{equation}\label{eqn:optimalOmegaMultiplexUniform}
	\omega=\frac{1}{T\langle\log\theta_\mathrm{in}-\log\theta_\mathrm{out}\rangle_t}\log\left( 1+\frac{p}{1-p}K\right)
\end{equation}
for uniform multiplex networks, so we see that $\omega_\mathrm{multiplex}=\omega_\mathrm{temporal}/T$. This scaling makes intuitive sense, as the number of directed interlayer edges per physical node is equal to $T-1$ for temporal networks and is equal to $T(T-1)$ for multiplex networks.


\section{Parameter estimation}\label{sec:parameterEstimation}

In Sec.~\ref{sec:multilayerEquiv}, we showed an equivalence between modularity maximization and statistical inference in SBMs, given certain assumptions, for different types of multilayer networks. In this section, we use these results to determine appropriate values for the resolution and interlayer coupling parameters that arise in the formulation of multilayer modularity. For concreteness, we focus on the uniform setting, in which there are only two parameters ($\gamma$ and $\omega$). However, one can straightforwardly extend all of our results in this section to the layer-dependent case.


\subsection{Iterative algorithm}\label{subsec:iterativeAlg.}

Throughout Sec.~\ref{sec:multilayerEquiv}, we derived expressions for $\gamma$ and $\omega$ that depend on the unknown quantities $\theta_\mathrm{in}$, $\theta_\mathrm{out}$, $p$, and $K$. We therefore have to estimate the values of all of these parameters. To do this, we use an iterative procedure similar to the one that was proposed in \cite{newman2016} for monolayer networks. Given some initial guesses $\gamma^{(0)}$ and $\omega^{(0)}$, the iterative process alternates between running modularity maximization with the current estimates of these parameters and using the resulting community structure to estimate $\theta_\mathrm{in}$, $\theta_\mathrm{out}$, $p$, and $K$, thereby obtaining new values for $\gamma$ and $\omega$. The algorithm stops once $\gamma$ and $\omega$ converge, up to some pre-specified tolerance. We give the pseudocode for this iterative procedure in Alg.~\ref{alg:itModMax}. The iteration proceeds similarly when the parameters are layer-dependent. 
It is also possible to use the framework of Alg.~\ref{alg:itModMax} to update only one of the two parameters, $\gamma$ and $\omega$, while keeping the other one fixed. A {\sc Matlab} implementation of Alg.~\ref{alg:itModMax} (which we call {\sc IterModMax}) is available at \url{https://github.com/roxpamfil/IterModMax}.

\begin{algorithm}[ht!]
	\small{
		\begin{algorithmic}
			\Function{IterativeModularityMaximization}{$\matt{A}$}
			\State \textbf{initialize} $\gamma=\gamma^{(0)}$ and $\omega=\omega^{(0)}$
			\While{not converged}
			\State $\vec{g}$ $\gets$ {\sc MaximizeModularity}$(\matt{A},\gamma,\omega)$  \Comment{Run multilayer modularity maximization (e.g., using {\sc GenLouvain} \cite{genlouvain})}
			\State $\theta_\mathrm{in},\theta_\mathrm{out},p,K \gets$ {\sc EstimateSBMParameters}$(\matt{A},\vec{g})$ 
			\\ \Comment{Use detected communities to estimate SBM parameters} 
			\State $\gamma \gets$ {\sc UpdateGamma}$(\theta_\mathrm{in},\theta_\mathrm{out})$ \Comment{Update $\gamma$ using Eqn.~\eqref{eqn:gammaOptimalUnif}}
			\State $\omega \gets$ {\sc UpdateOmega}$(\theta_\mathrm{in},\theta_\mathrm{out},p,K)$  \Comment{Update $\omega$ using Eqn.~\eqref{eqn:omegaOptimalTemporal} or Eqn.~\eqref{eqn:optimalOmegaMultiplexUniform}}
			\EndWhile
			\State \textbf{return} $\gamma$, $\omega$ \Comment{Optimal modularity parameters}
			\State \textbf{return} $\vec{g}$ \Comment{Detected communities using optimal parameters}
			\EndFunction
		\end{algorithmic}
	}
	\caption{Iterative algorithm for performing modularity maximization and estimating resolution and interlayer-coupling parameters in a multilayer network.}
	\label{alg:itModMax}
\end{algorithm}

A key step of Alg.~\ref{alg:itModMax} is estimating $\theta_\mathrm{in}$, $\theta_\mathrm{out}$, $p$, and $K$, given a multilayer network's supra-adjacency matrix $\matt{A}$ and the current multilayer partition $\vec{g}$. This task is relatively straightforward for temporal and multilevel networks, although it is more involved for multiplex networks. We explain how to perform this estimation in Sec. \ref{sm:sec:estimateSBM} of our supplementary materials.


\subsection{Numerical examples}\label{subsec:numericalExamples}

\begin{figure}[ht]
	\centering
	\subfloat[$\gamma=1$ and $\omega=1$]{\includegraphics[width=0.3\textwidth]{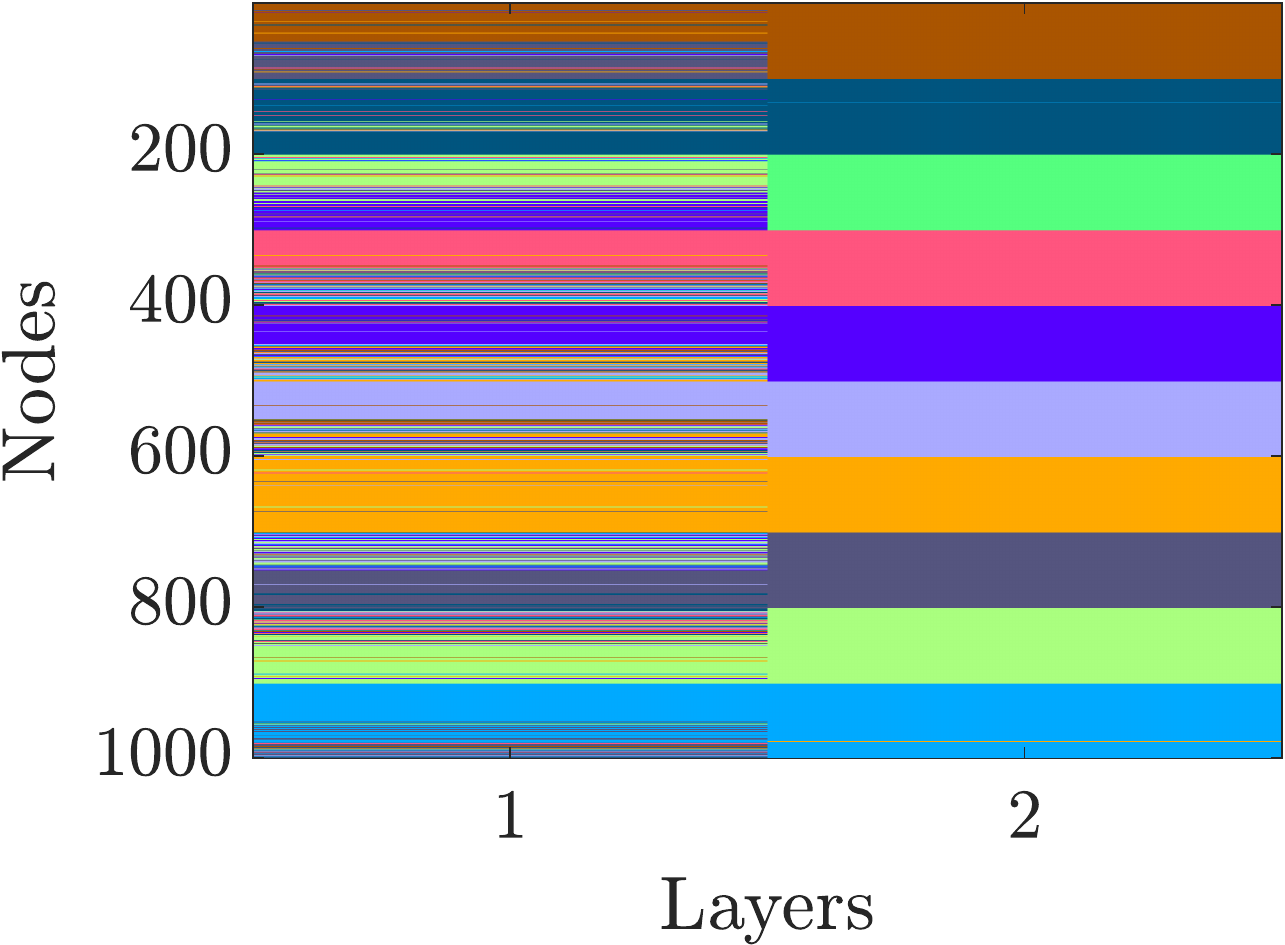}} \hspace{6em}
	\subfloat[$\gamma \approx 1.60$ and $\omega \approx 1.30$]{\includegraphics[width=0.3\textwidth]{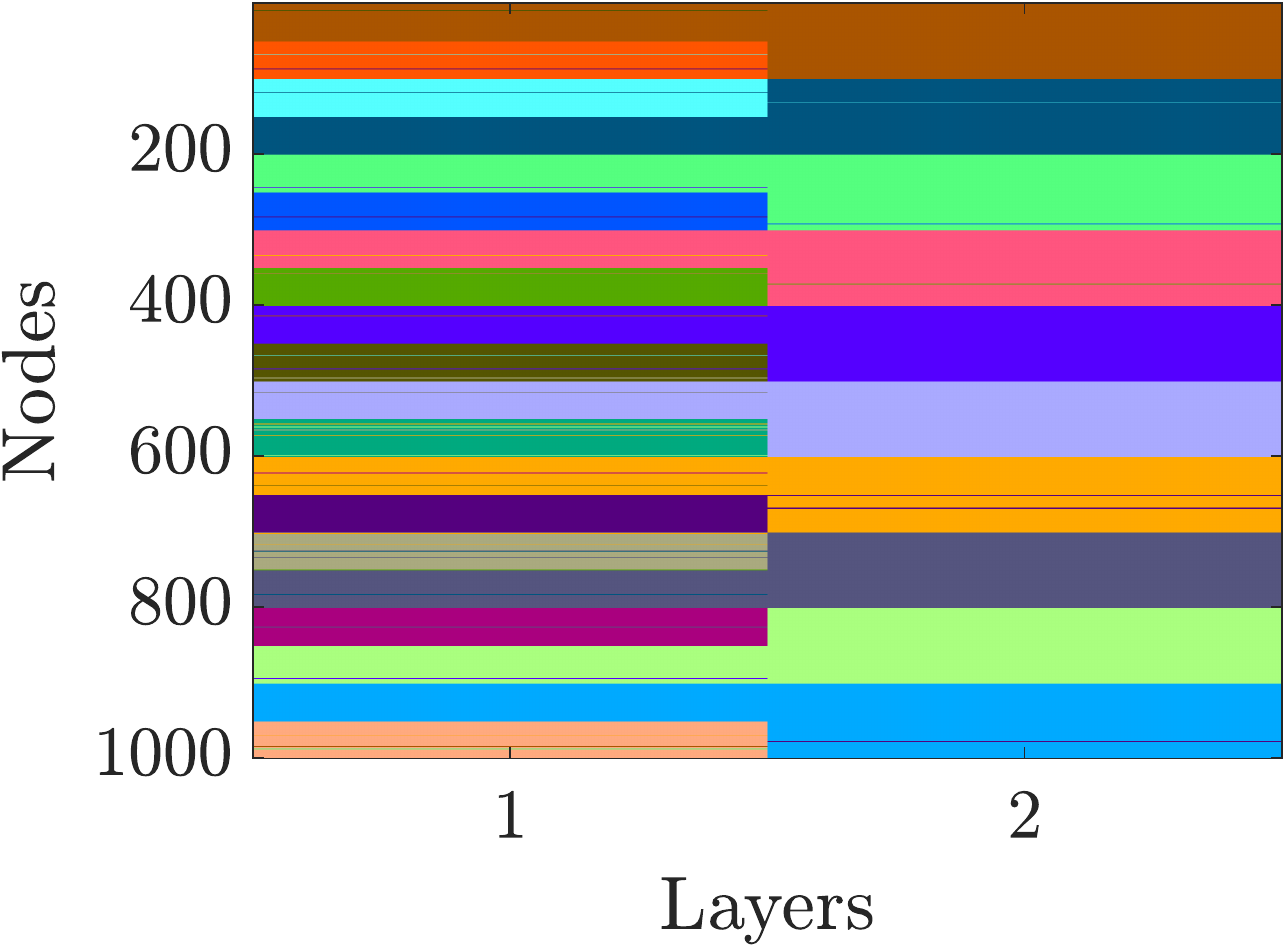}}
	\caption{Detected multilayer partitions for a toy network with 20 communities in the first layer that merge pairwise into 10 communities in the second layer. We set the values of $\gamma$ and $\omega$ to (a) $\gamma=1$ (the ``default'' value) and $\omega=1$ and (b) the optimal values $\gamma \approx 1.60$ and $\omega \approx 1.30$ that we infer using Alg. \ref{alg:itModMax}. In both cases, the normalized mutual information (NMI) between the detected partitions and the planted one is close to $1$ for the second layer. For the first layer, the NMI is approximately $0.5$ using the naive approach, and it is about $0.95$ for the optimized approach.}
	\label{fig:simpleExample}
\end{figure}

We first illustrate the performance of Alg.~\ref{alg:itModMax} on a simple network. Consider a network with $N=1000$ nodes, $T=2$ layers, and $K_1=20$ communities in the first layer that merge pairwise into $K_2=10$ communities in the second layer. We place edges independently in the two layers with probabilities $p_\mathrm{in}=0.32$ (for nodes in the same community) and $p_\mathrm{out}=0.1$ (for nodes in different communities). We choose these values near the detectability threshold for a monolayer network to make the community-detection task sufficiently difficult. Note that the parameters $p_\mathrm{in}$ and $p_\mathrm{out}$ are not the same as the parameters $\theta_\mathrm{in}$ and $\theta_\mathrm{out}$ from our generative model, as the latter are also multiplied by node degrees to generate edge probabilities. In Fig.~\ref{fig:simpleExample}, we show the detected communities for two parameter choices: (1) the parameter values $\gamma=1$ and $\omega=1$ and (2) the values $\gamma \approx 1.60$ and $\omega \approx 1.30$ to which the iteration from Alg.~\ref{alg:itModMax} converges. The naive approach recovers the correct structure in the second layer but not in the first; in particular, it identifies $10$, rather than $20$, communities in the first layer. By contrast, our iterative modularity-maximization algorithm identifies a multilayer partition that is close to the planted one.

We also test our parameter-estimation approach on existing multilayer network benchmarks, in which one can easily tune the strength of planted community structure by changing one parameter. This flexibility allows us to study the community-detection problem at various levels of difficulty in a systematic way. In this section, we provide results for the generative temporal network model that was proposed by Ghasemian et al. \cite{ghasemian2016}. In Sec. \ref{sm:sec:numericalExamples} of the supplementary materials, we test our approach on temporal and multiplex networks generated using the models of Bazzi et al. \cite{bazzi2016generative}.

Apart from the lack of degree correction, the model of Ghasemian et al. \cite{ghasemian2016} is the same as the one that we used in Sec.~\ref{subsec:temporal} to study multilayer representations of temporal networks. In particular, it uses the same copying and resampling process to propagate community labels across temporal layers. In any layer other than the first, a node keeps its community label from the previous layer with probability $\eta$, and it samples a label uniformly at random (from $K$ available labels) with probability $1-\eta$. The parameter $\eta$ is the same as our copying probability $p$, but we use different notation to emphasize that one is a tuning parameter for generating a network and the other is a parameter that we infer while detecting communities in that network. 
The expected community sizes are equal to $N/K$ for all communities and all layers. 

After assigning nodes to communities in this way, we place edges independently in each layer with probabilities $p_\mathrm{in}$ (if the nodes are in the same community) and $p_\mathrm{out}$ (if the nodes are in different communities). Because this generative model has no degree correction, it produces networks in which each node has the same expected degree in every layer. The ratio $\epsilon=p_\mathrm{out}/p_\mathrm{in}$ controls the strength of the planted community structure.

\begin{figure*}[htbp]
	\centering
	\subfloat[{\sc GenLouvain}]{\includegraphics[width=0.45\textwidth]{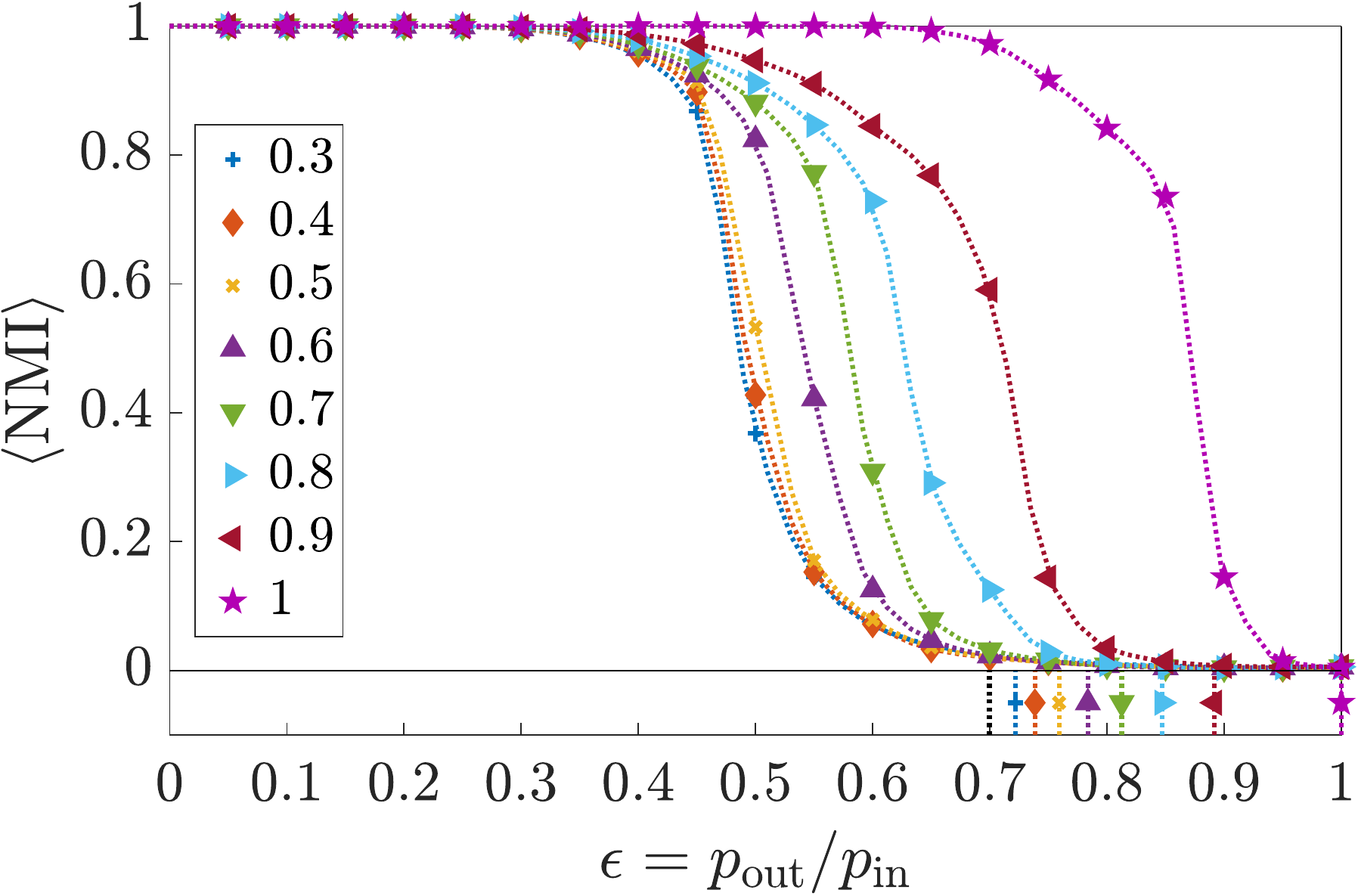}} \hfill
	\subfloat[{\sc GenLouvainRand}]{\includegraphics[width=0.45\textwidth]{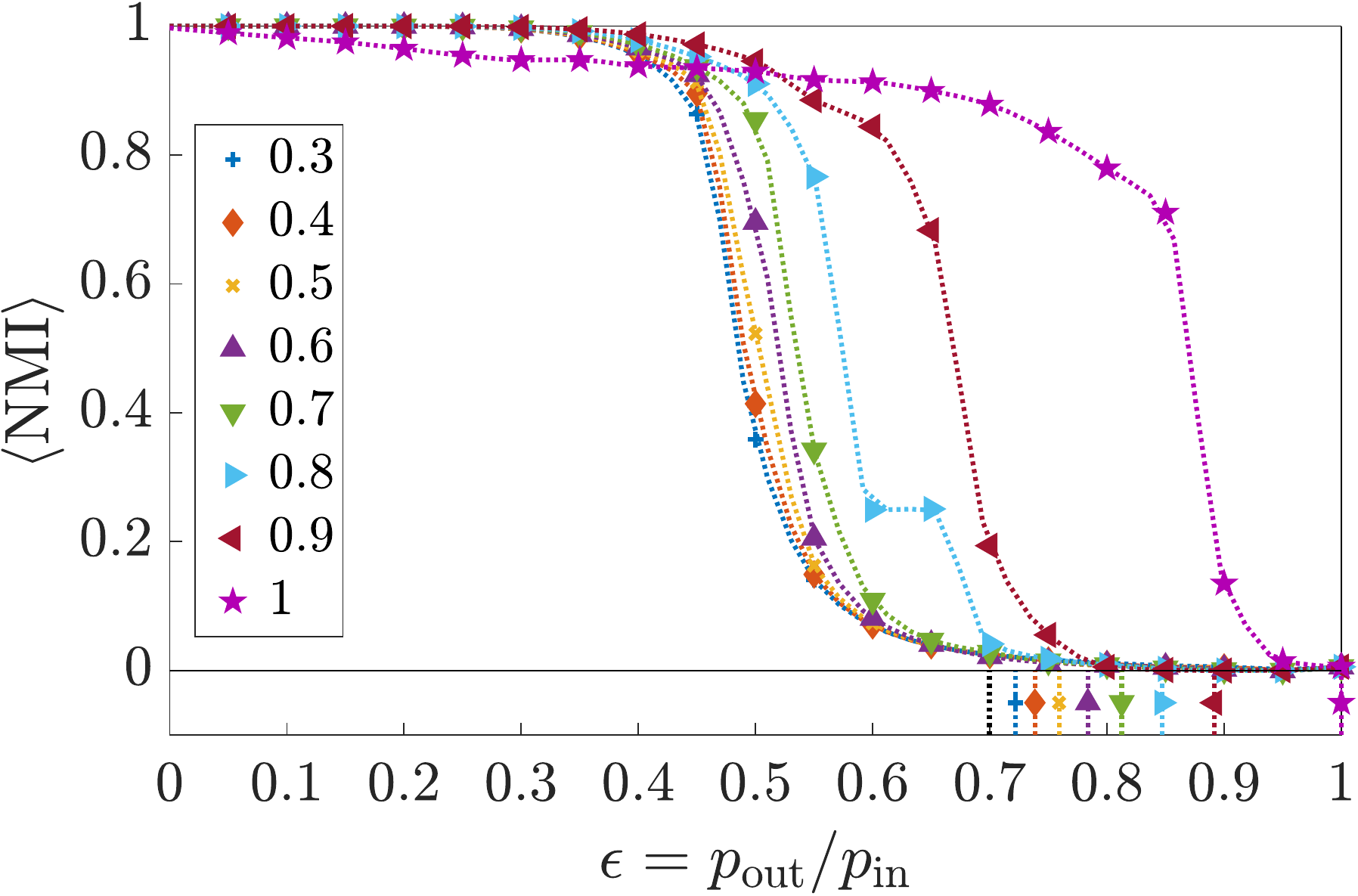}}
	\caption{Results for the temporal multilayer benchmark network from \cite{ghasemian2016}. The plots show layer-averaged NMI scores between the planted partition and the one detected by modularity maximization as a function of community strength $\epsilon$ for two different versions of the {\sc GenLouvain} algorithm \cite{genlouvain}. Each line and set of markers corresponds to a different value of the copying probability $\eta$, and each data point is a mean over $100$ trials. We use the theoretically-optimal $\gamma$ and $\omega$ for each choice of parameters. The bottom parts of the plots (vertical lines) indicate the detectability thresholds from \cite{ghasemian2016}, beyond which no algorithm is able to detect the planted structure. The plots differ in how the {\sc GenLouvain} algorithm explores the space of multilayer partitions to increase modularity: (a) choose the move that yields the largest increase in modularity; (b) choose a move with a probability proportional to the resulting increase in modularity. The algorithm in panel (a) seems to perform better, especially for large $\eta$, although we note that this version of the algorithm can lead to poor behavior in certain cases and for sufficiently large $\omega$ \cite{bazzi2016}.}
	\label{fig:dynamicModel}
\end{figure*}

For their numerical experiments in \cite{ghasemian2016}, Ghasemian et al. limited themselves to situations
with $K=2$ communities, and we do the same in the present paper. We construct networks with $T=40$ layers and $N=512$ nodes in each layer. Each node has a mean degree of $c=32$, and we vary $\eta$ and $\epsilon$ in our calculations. In qualitative terms, the ability of a community-detection algorithm to recover a planted partition should decrease as $\epsilon$ increases (because community structure is becoming weaker), and it should increase as $\eta$ increases (because the multilayer partition is changing less between successive layers). We show the results of these tests in Fig.~\ref{fig:dynamicModel}. Note that our plots use the theoretically-optimal values of $\gamma$ and $\omega$ from Eqns.~\eqref{eqn:gammaOptimalTemporal} and \eqref{eqn:omegaOptimalTemporal}, rather than iterating from some initial values. We choose to proceed in this manner because the authors of \cite{ghasemian2016} also assumed that all parameters aside from the community assignments $\vec{g}$ are known. Towards the bottom of each plot, we also show the detectability thresholds derived in \cite{ghasemian2016}; beyond these values, no algorithm can recover the planted communities. These thresholds, which depend on $\eta$, 
are derived in the limit in which the number $T$ of layers tends to infinity. 
We measure the level of success of each run by averaging over all layers the normalized mutual information \footnote{We use the version of NMI in which one normalizes by the mean entropy of the two input partitions \cite{li2001nmi}.} (NMI) between the planted partition $g_\mathrm{true}$ and the algorithmically-detected one:
\begin{equation*}
	\left\langle \mathrm{NMI}(\vec{g},\vec{g}_\mathrm{true}) \right\rangle=\frac{1}{T}{\sum_{t=1}^T{\mathrm{NMI}(\vec{g}^t,\vec{g}^t_\mathrm{true})}}\,.
\end{equation*}

The two different plots in Fig.~\ref{fig:dynamicModel} correspond to two different versions
of the {\sc GenLouvain} algorithm for optimizing multilayer modularity
\cite{genlouvain}. When the {\sc GenLouvain} algorithms seek to increase the value of modularity by reassigning nodes to communities, one can either (1) select the move that leads to the largest increase in modularity ({\sc GenLouvain}) or (2) use a ``weighted" random move, which is chosen with a probability proportional to the resulting increase in modularity ({\sc GenLouvainRand}). Previous work \cite{bazzi2016} found that the original version of the {\sc GenLouvain} algorithm can sometimes lead to undesirable behavior in which interlayer modularity terms are overemphasized relative to intralayer terms for sufficiently large $\omega$. The {\sc GenLouvainRand} version of the algorithm was designed to avoid this problem. For our numerical experiments in Fig.~\ref{fig:dynamicModel}, we find that {\sc GenLouvain} tends to perform better. This is due to two factors. First, at small values of $\eta$, the theoretically-optimal values of $\omega$ do not fall in the problematic regime that was described in \cite{bazzi2016}. Second, at large values of $\eta$, the fact that {\sc GenLouvain} overemphasizes persistence tends to help, rather than hinder, the algorithm's performance. 
Note that the results in Fig.~\ref{fig:dynamicModel} do not contradict the earlier findings from \cite{bazzi2016}.

Our results in Fig.~\ref{fig:dynamicModel} exhibit the same qualitative features as similar plots in the paper that introduced these benchmarks \cite{ghasemian2016}. Specifically, for each value of $\eta$, the success rate of the algorithms (as measured by NMI) is close to $1$ for small values of $\epsilon$, and it then drops rapidly to $0$ near some critical value of $\epsilon$ that depends on $\eta$. This degradation in performance occurs before the theoretically-predicted detectability thresholds that were derived in \cite{ghasemian2016}. By contrast, the belief-propagation algorithm proposed in 
\cite{ghasemian2016} performs well
closer to these thresholds. There are several reasons for this observation. The first is that belief-propagation methods are known to give asymptotically optimal accuracy, in that they are able to recover planted communities all the way down to the threshold, as long as there is sufficient data (in terms of network size and number of layers) and as long as the relevant parameters are initialized sufficiently close to their optimal values \cite{decelle2011a,decelle2011b,moore2017}. 
By contrast, there are no such theoretical guarantees for modularity maximization. Additionally, the belief-propagation algorithm in \cite{ghasemian2016} only looks for multilayer partitions with two communities, whereas our modularity maximization algorithm also considers other partitions.
Another possible reason for the suboptimal results in Fig.~\ref{fig:dynamicModel} may be that our values of $\gamma$ and $\omega$,
which we calculated from Eqns.~\eqref{eqn:gammaOptimalTemporal} and \eqref{eqn:omegaOptimalTemporal}, are not the best possible values to use in practice. We investigate this next.

\begin{figure*}[htbp]
	\centering
	\subfloat[``Easy" regime ($\eta=0.7$, $\epsilon=0.4$)]{\includegraphics[width=0.35\textwidth]{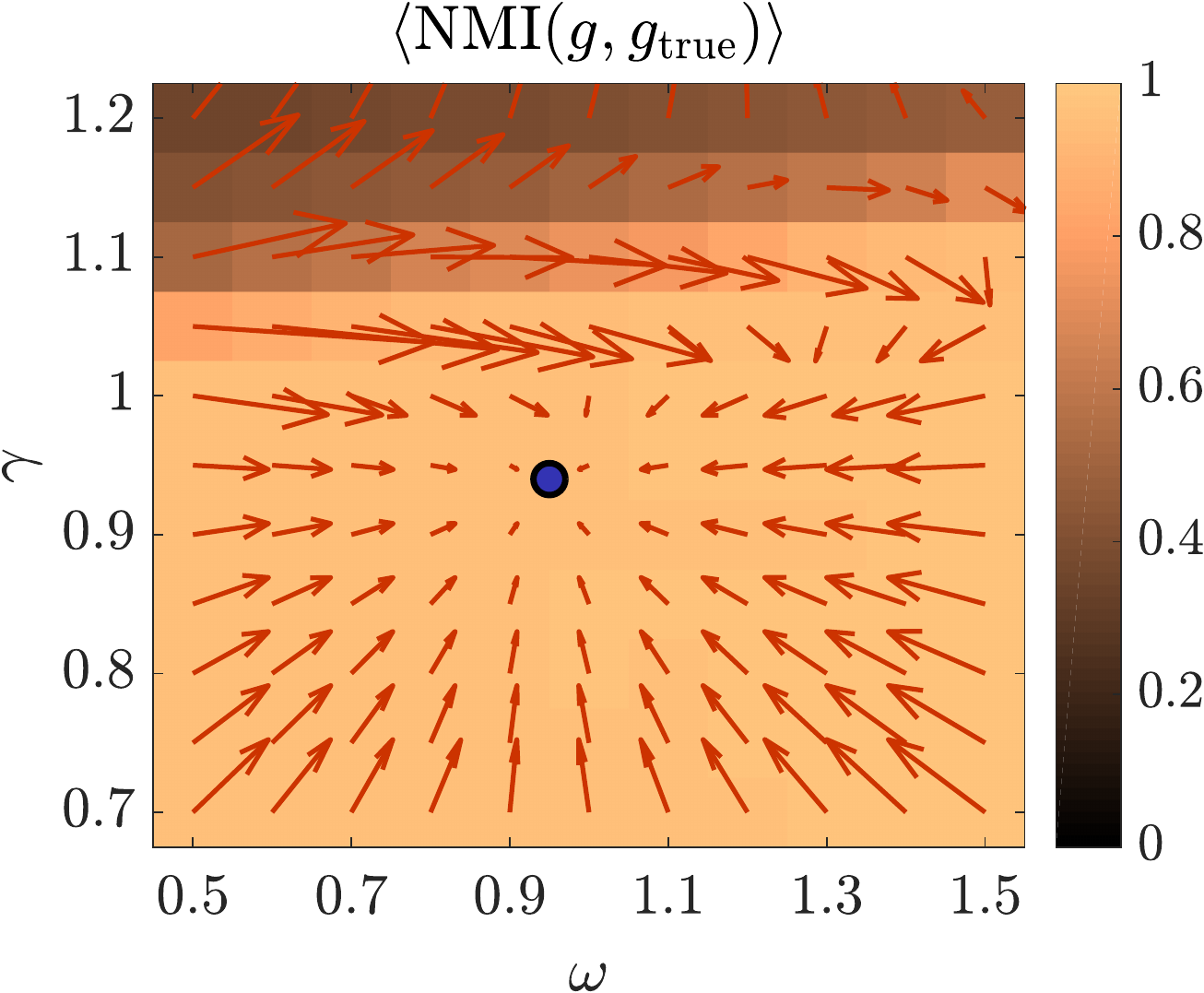}} \hspace{0.1\textwidth}
	\subfloat[``Hard" regime ($\eta=0.5$, $\epsilon=0.5$)]{\includegraphics[width=0.35\textwidth]{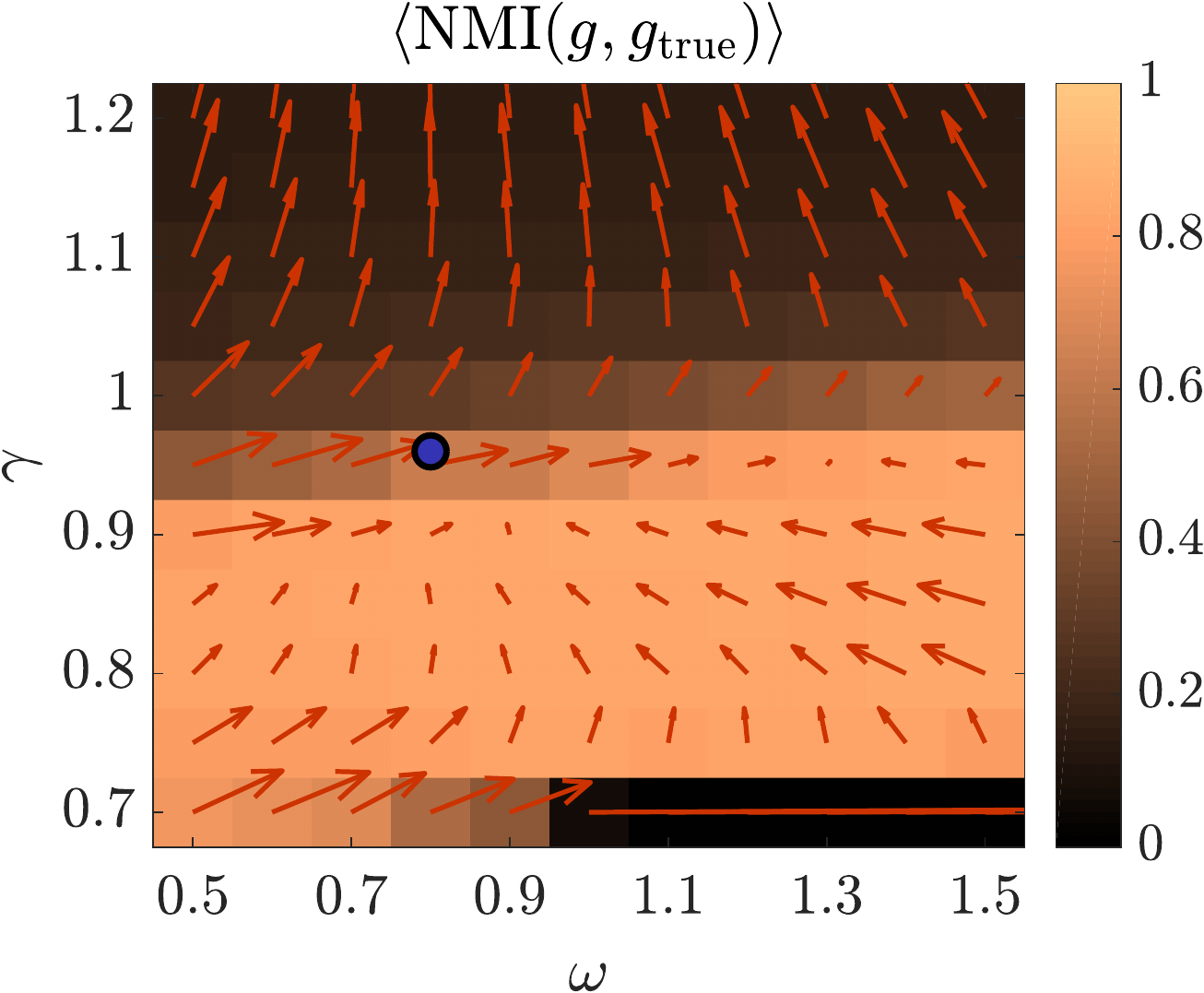}}
	\caption{Illustration of the dynamics of the iteration in Alg.~\ref{alg:itModMax}. In each plot, the heat map shows NMI scores between the planted partition $\vec{g}_\mathrm{true}$ and the algorithmically-detected partition $\vec{g}$, averaged over $10$ runs for each combination of $\gamma$ and $\omega$ values. The arrows centered at each $(\gamma,\omega)$ grid cell indicate the direction of the new $\gamma$ and $\omega$ estimates (averaged over the $10$ trials) that we obtain from Eqns.~\eqref{eqn:gammaOptimalTemporal}--\eqref{eqn:omegaOptimalTemporal} after performing modularity maximization. For clarity, we have scaled down the arrow sizes. The blue disk in each plot indicates the location of the correct parameter values (namely, $\gamma_\mathrm{opt} \approx 0.94$ and $\omega_\mathrm{opt} \approx 0.95$), which one can calculate from the planted partition.
		(a) Example with $\eta=0.7$ and $\epsilon=0.4$ (an ``easy" case). As the arrows indicate, the iteration converges to the true parameter values for many initial values $\gamma^{(0)}$ and $\omega^{(0)}$. (b) Example with $\eta=0.5$ and $\epsilon=0.5$ (a ``hard" case). The region in the plane that results in the highest NMI values is roughly
		$0.8 \lessapprox \gamma \lessapprox 0.9$ and $0.7 \lessapprox \omega \lessapprox 1.5$. The optimal parameter values $\gamma_\mathrm{opt} \approx 0.96$ and $\omega_\mathrm{opt} \approx 0.80$ lie slightly outside 
		this regime. Although many of the arrows point in the direction of the blue disk, once the iteration continues from
		there, 
		it results in increasingly large resolution values and thus never converges. However, by returning the largest-modularity partition that it encounters during the iterative process, the algorithm identifies a solution that is close to the planted structure. (The NMI is about $0.86$.)}
	\label{fig:heatmapVectorField}
\end{figure*}

In Fig.~\ref{fig:heatmapVectorField}, we show heat maps of mean NMI values for two different choices of $\eta$ and $\epsilon$. Specifically, after generating a multilayer network for each of the two settings, we sample values of $\gamma$ and $\omega$ from a two-dimensional grid and run {\sc GenLouvainRand} $10$ times for each pair of values. The mean NMI values in each grid cell indicate the level of agreement between the true partition and the one that we detect algorithmically. Figure~\ref{fig:heatmapVectorField}(a) corresponds to an ``easy" case, in which modularity maximization correctly recovers the planted partition at the optimal parameter values; and Fig.~\ref{fig:heatmapVectorField}(a) corresponds to a ``hard" case, in which we are only partially able to recover the planted structure. (Situations in which there is no overlap between the detected partition and the planted one are not particularly interesting.) In each of these plots, we mark the optimal parameter values $\gamma_\mathrm{opt}$ and $\omega_\mathrm{opt}$ with a blue disk. The arrows centered at each $(\gamma,\omega)$ grid point indicate the direction of the new $\gamma$ and $\omega$ estimates (averaged over the $10$ trials) that we obtain from Eqns.~\eqref{eqn:gammaOptimalTemporal}--\eqref{eqn:omegaOptimalTemporal} after performing modularity maximization. (For clarity, we have scaled down the arrow sizes.) This visualization indicates how our iterative approach explores (on average) the two-dimensional parameter space from different starting positions. In Fig.~\ref{fig:heatmapVectorField}(a), most of the arrows point to the location of the optimal point $(\gamma_\mathrm{opt},\omega_\mathrm{opt})$. We are thus in the ``easy" regime, in which the iteration from Alg.~\ref{alg:itModMax} converges to the correct parameter values and recovers the correct partition. The situation is more complicated in Fig.~\ref{fig:heatmapVectorField}(b). We observe that the optimal parameter values lie slightly outside the region in the $(\gamma,\omega)$ plane with
the largest mean NMI values. Although many of the arrows point in the direction of the optimal point $(\gamma_\mathrm{opt},\omega_\mathrm{opt})$, {\sc GenLouvainRand} detects suboptimal partitions into communities at these values. (We confirmed that the detected partitions are indeed suboptimal, as they give a smaller modularity value than the planted partition.) The arrows also illustrate why the iterative approach fails to converge in this case. From the optimal point, the resolution value becomes increasingly large, with no hope of returning to the large-NMI regime that is roughly in the region
$0.7 \lessapprox \gamma \lessapprox 0.9$. This observation suggests that there is no stable fixed point of the iteration in Fig.~\ref{fig:heatmapVectorField}, although there may be
an unstable fixed point.

To circumvent the issue of the resolution $\gamma$ increasing indefinitely for some networks (which tend to have relatively weak community structure), we find it
useful to decrease $\gamma$ in the iterations of Algorithm~\ref{alg:itModMax} whenever the number of communities exceeds some user-specified threshold. Specifically, we decrease $\gamma$ by $20\%$ (leaving $\omega$ unchanged) whenever $K$ exceeds some given $K_\mathrm{max}$. This helps our algorithm explore the space of resolution and interlayer-coupling parameters in a useful way, even when it fails to converge to a fixed point. The algorithm can then return the largest-modularity partition that it encounters during this exploration. For the multilayer network in Figure~\ref{fig:heatmapVectorField}(b), this approach finds a solution with an NMI score of $0.86$. This example illustrates an important point: even when our proposed iterative algorithm does not converge to a fixed point, it can still identify meaningful community structure in a multilayer network. An important topic for future work is to devise better heuristics (or alternatives to the fixed-point iteration of Algorithm \ref{alg:itModMax}) for this parameter-estimation problem.


\section{Application: Lazega Law Firm}\label{sec:lazega}

We illustrate our approach on the Lazega Law Firm network \cite{lazega2001}. This multiplex network encompasses interactions between $N=71$ partners and associates who work at the same law firm. The network has $T=3$ layers, each with directed edges, which encode \textit{co-work}, \textit{friendship}, and \textit{advice} relationships. The data set also includes 7 pieces of metadata, which we can use to analyze detected communities. These are as follows: (1) status (partner or associate); (2) gender; (3) office (Boston, Hartford, or Providence); (4) seniority (specifically, years with the firm); (5) age; (6) practice (litigation or corporate); (7) law school (Harvard, Yale, University of Connecticut, or other). We group both the seniority and age metadata into 5-year bins.

\begin{figure}[htb!]
	\centering
	\subfloat[Scatter plot of fixed points]{\includegraphics[width=0.35\textwidth]{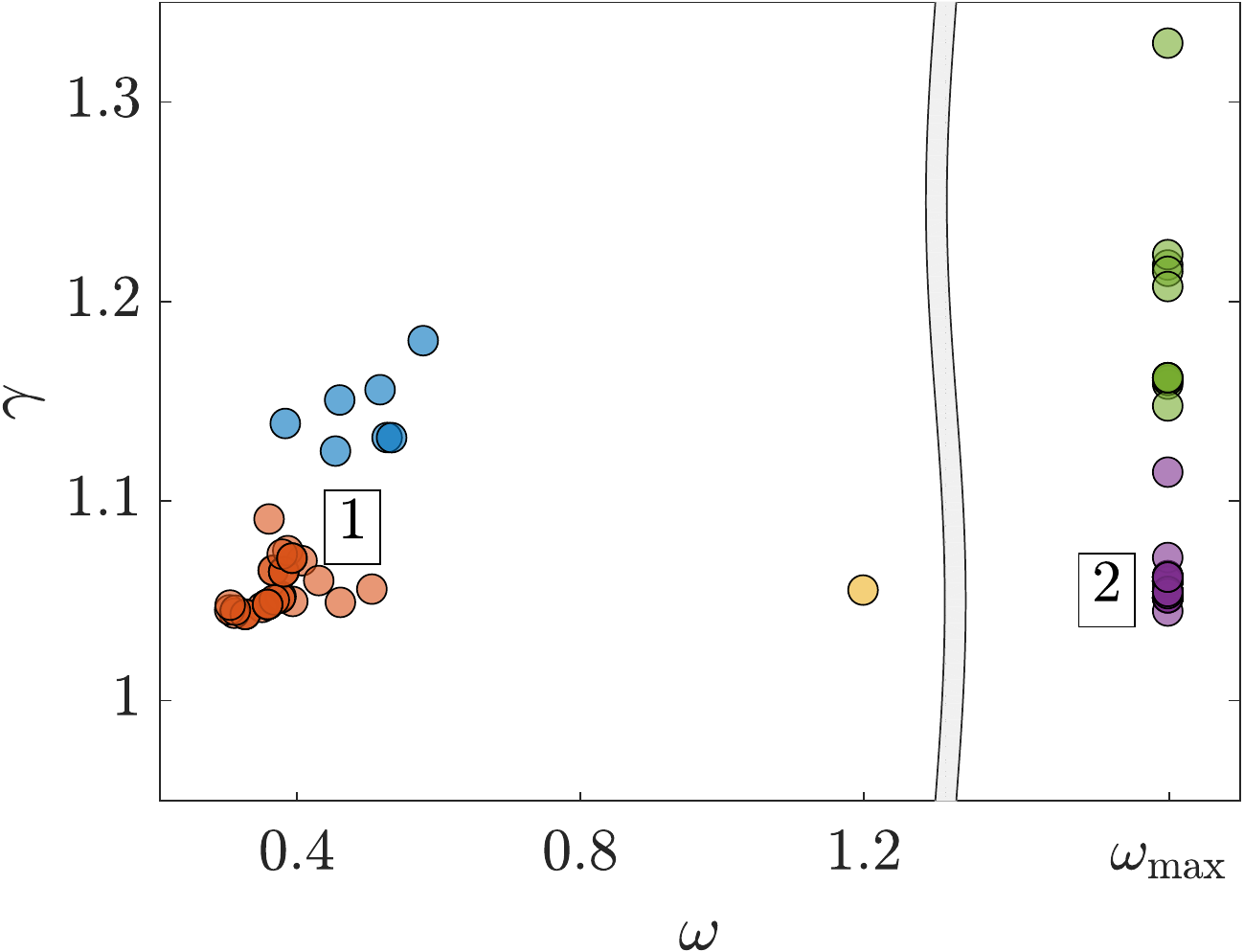}} \hspace{4em}
	\subfloat[Matrix of NMI scores]{\includegraphics[width=0.34\textwidth]{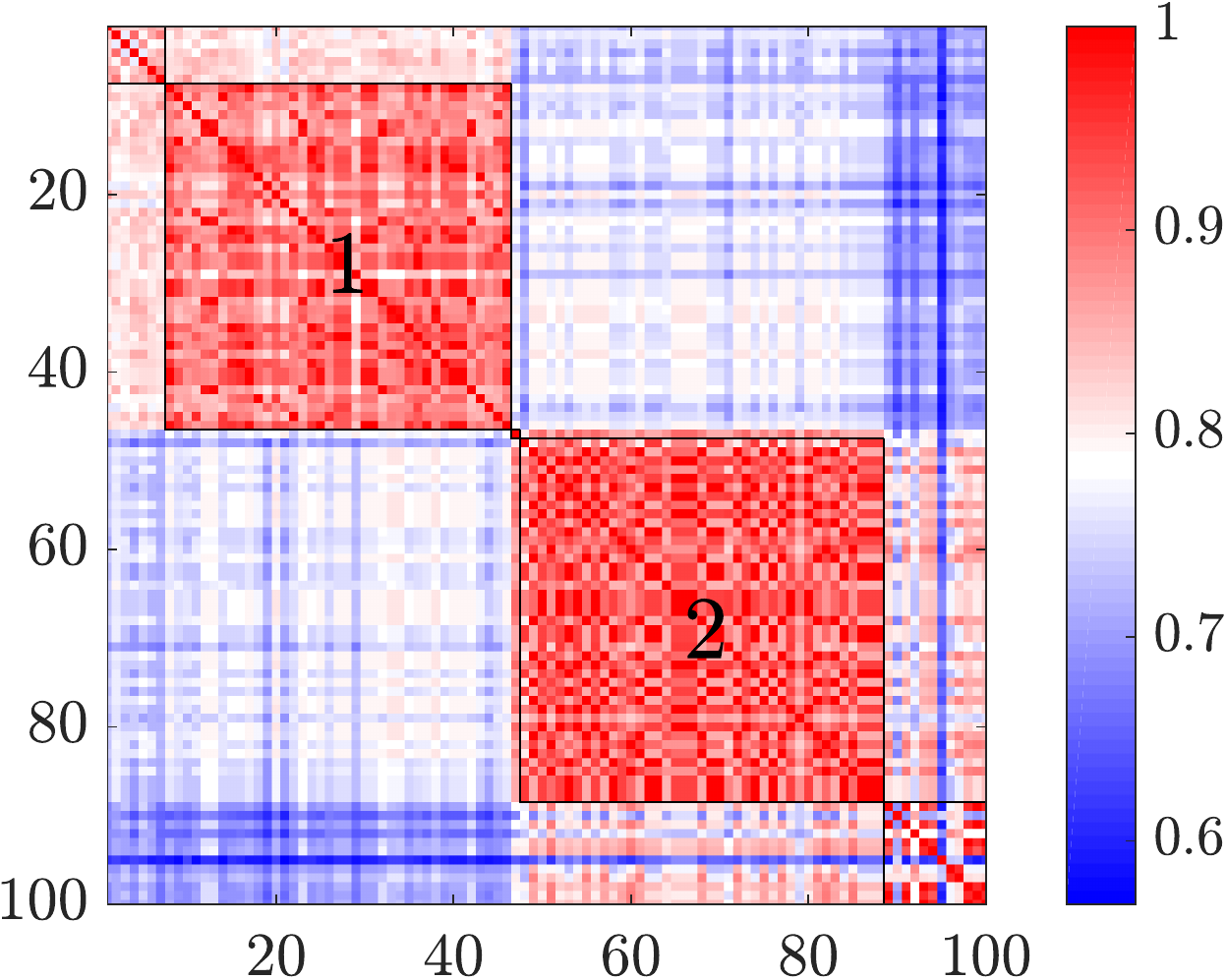}}
	\caption{(a) Scatter plot of $(\gamma_\mathrm{opt},\omega_\mathrm{opt})$ fixed points that our iterated modularity-maximization algorithm identified when we apply it $100$ times to the Lazega Law Firm network. The colors indicate membership to one of $5$ groups that we identify using the $K$-means clustering algorithm \cite{jain1999}. About $80$ runs converge to one of the groups labeled $1$ and $2$.
		(b) Matrix of layer-averaged NMI values between all pairs of partitions that we identify across $100$ trials of our iterated modularity-maximization algorithm. The $5$ diagonal blocks correspond to the $5$ groups from panel (a).}
	\label{fig:lazegaScatterNMI}
\end{figure}

We apply a version of Alg.~\ref{alg:itModMax} for directed networks (see Sec.~\ref{sm:sec:directed} in supplementary materials for a derivation of this case) to the Lazega Law Firm data set, and we use {\sc GenLouvainRand} for community detection. (Using {\sc GenLouvain} gives very similar results.) Specifically, we run $100$ trials of our iterative algorithm using initial values of $\gamma$ and $\omega$ that we distribute uniformly in the intervals $[0,5]$ and $[0,1]$, respectively. Each trial converges within $30$ iterations (which is our specified maximum), and typically our procedure reaches a fixed point in a much smaller number of iterations. In Fig.~\ref{fig:lazegaScatterNMI}(a), we show a scatter plot of the resulting fixed points $(\gamma_\mathrm{opt},\omega_\mathrm{opt})$. This plot has two distinct regions of interest, depending on whether the copying probability $p$ that our algorithm learns
is strictly smaller than $1$ or is equal to $1$. In the former case, there is some variation in community structure across layers. In the latter case, communities are identical across all $3$ layers, and the theoretically-optimal coupling value is $\omega_\mathrm{opt}=\infty$. In the algorithm, we set $\omega$ equal to $\omega_\mathrm{max}=1000$ to introduce a very large penalty to nodes that switch communities across layers. 
We apply the $K$-means clustering algorithm to the (suitably normalized) points in Figure~\ref{fig:lazegaScatterNMI}(a) to assign them to $5$ groups, of which $3$ correspond to the runs with $p<1$ and the other $2$ correspond to the runs with $p=1$. Of the $100$ runs, about $80$ converge to one of the groups labeled $1$ and $2$ in Fig.~\ref{fig:lazegaScatterNMI}(a). To compare the outputs of different trials, we calculate the layer-averaged NMI between all pairs of partitions that we obtain across our $100$ trials, and we show the resulting similarity matrix in Figure~\ref{fig:lazegaScatterNMI}(b). The black borders around the diagonal blocks of this similarity matrix correspond to the $5$ groups from Figure~\ref{fig:lazegaScatterNMI}(a). We observe large similarity scores among trials in the same group, especially for the two dominant groups (labeled $1$ and $2$). Because one of them has $p<1$ and the other has $p=1$, we analyze them separately. 

\begin{figure}[htb!]
	\centering
	\subfloat[Consensus partition for cluster 1]{\includegraphics[width=0.33\textwidth]{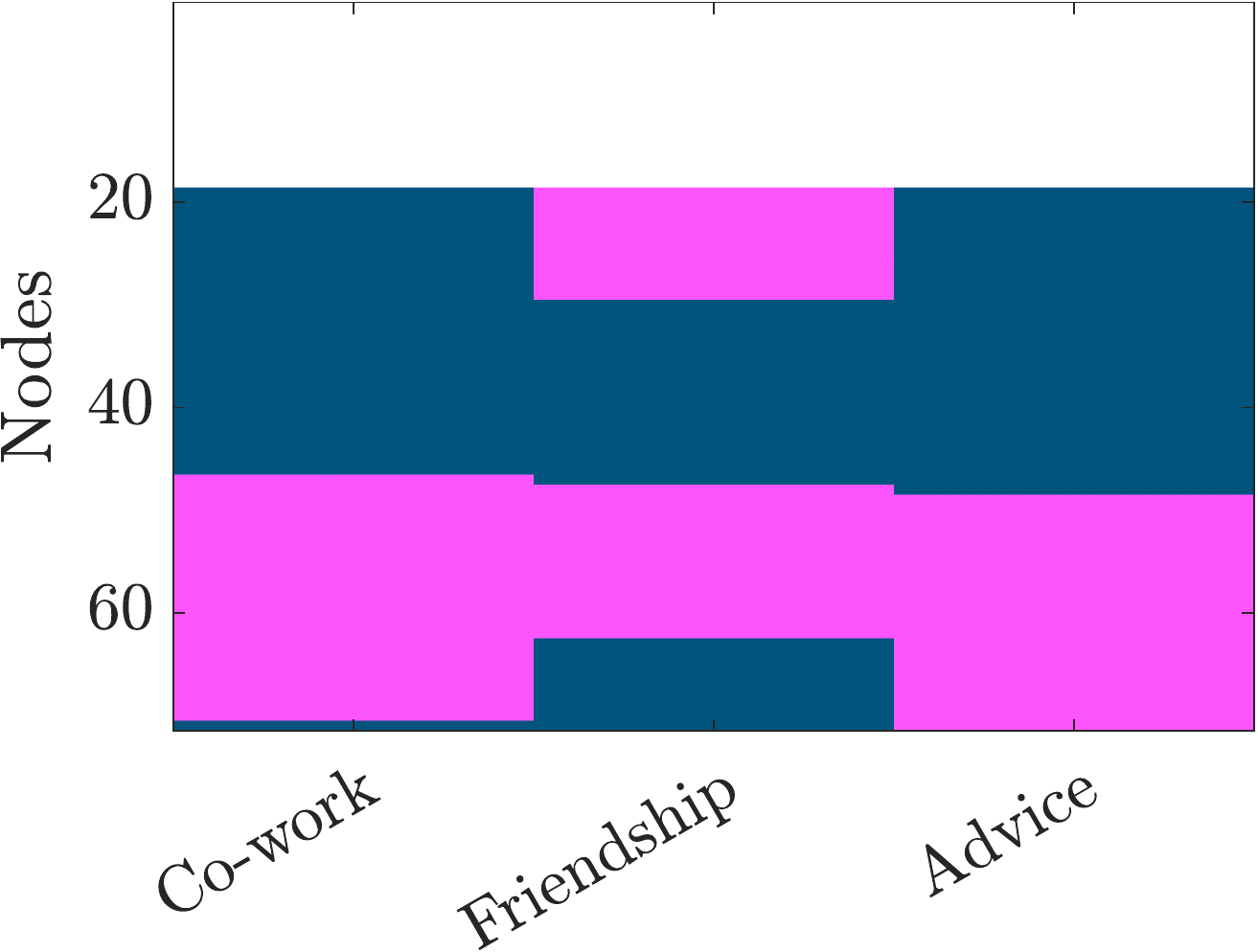}} \hspace{0.1\textwidth}
	\subfloat[Consensus partition for cluster 2]{\includegraphics[width=0.33\textwidth]{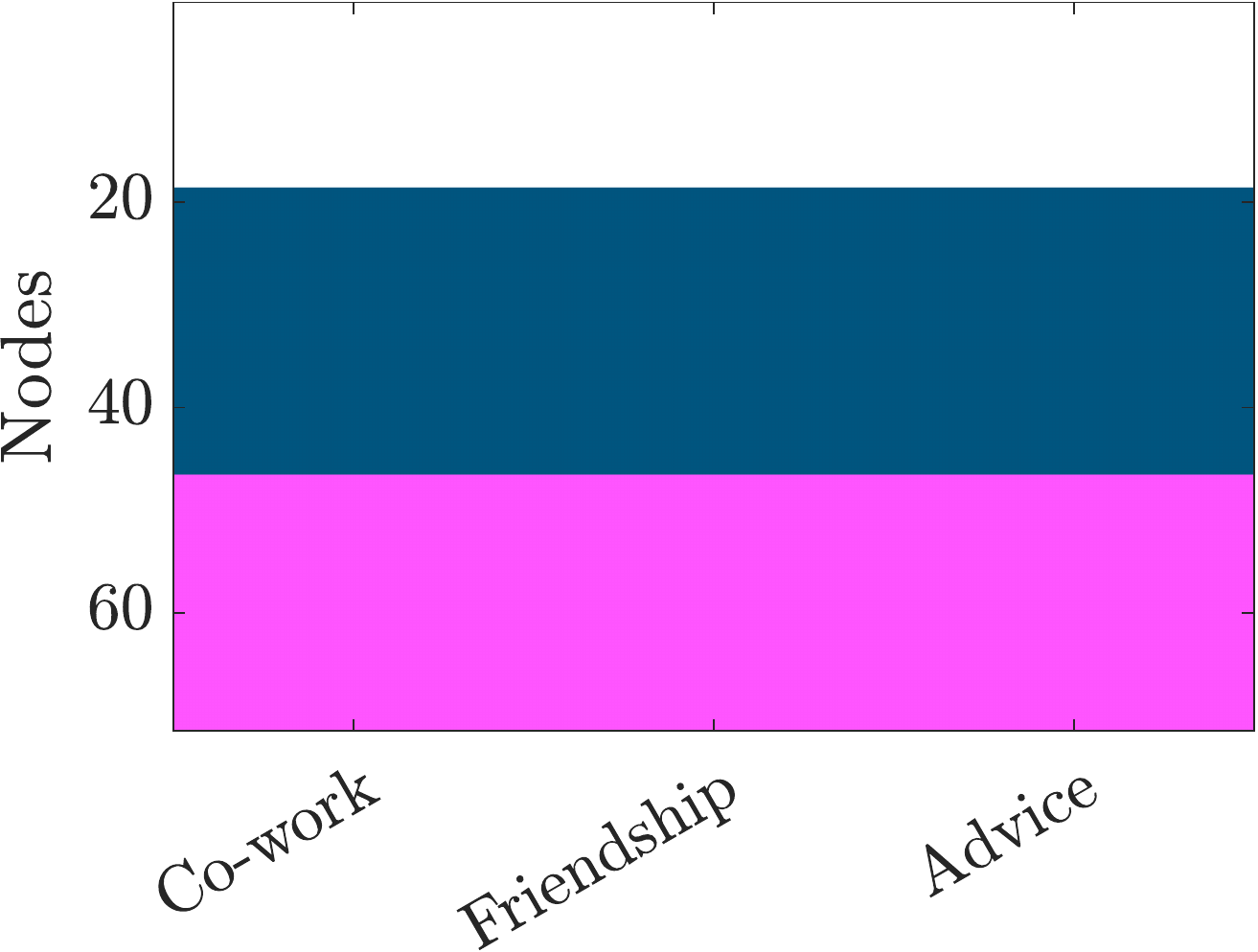}}
	\caption{Visualizations of the consensus partitions that correspond to the two largest groups from Fig.~\ref{fig:lazegaScatterNMI}(a). We use the same ordering of the rows in both panels. The partition in (a) reveals a set of nodes that are in a different community in the friendship layer than they are in the other two layers. The partition in (b) has identical communities in the three layers.}
	\label{fig:lazegaConsensus}
\end{figure}

\begin{figure}[htb!]
	\centering
	\subfloat[Cluster 1]{\includegraphics[width=0.27\textwidth]{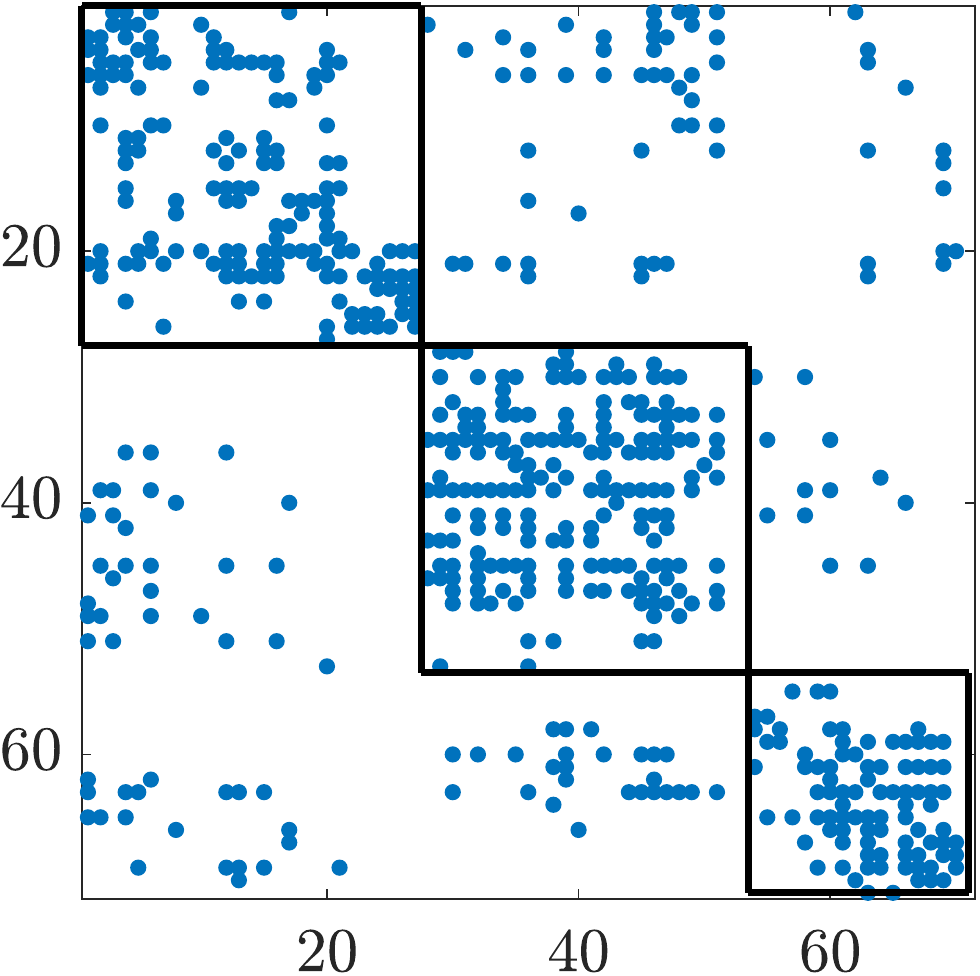}} \hspace{0.15\textwidth}
	\subfloat[Cluster 2]{\includegraphics[width=0.27\textwidth]{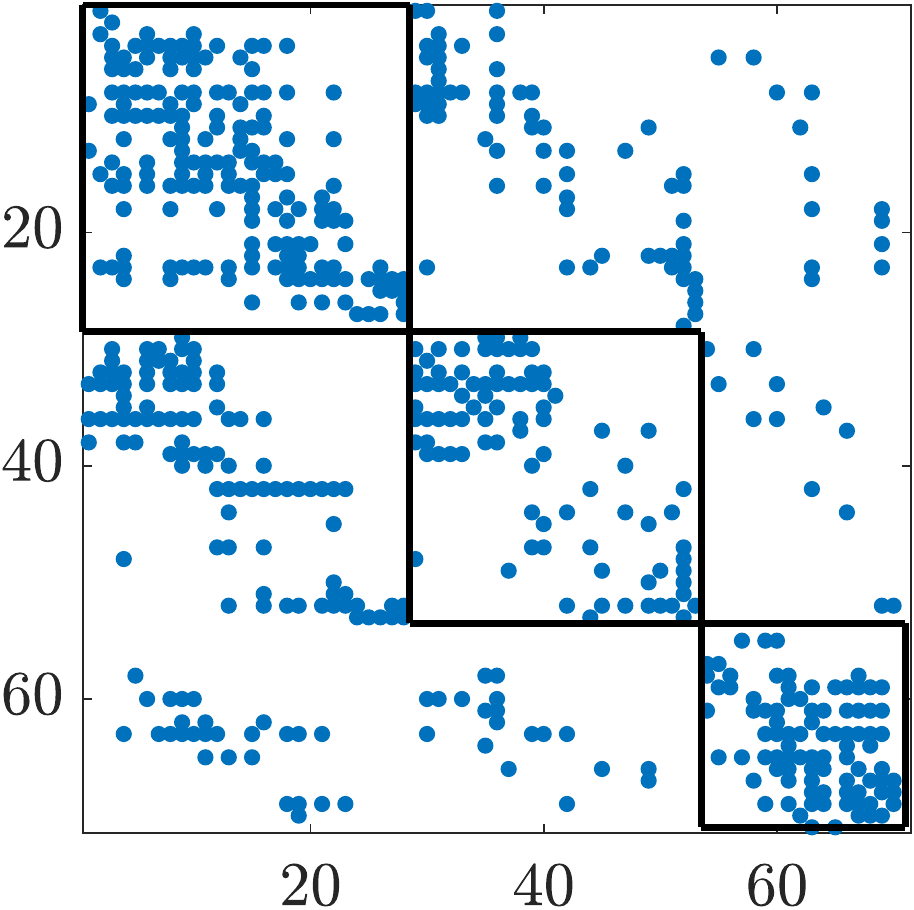}} 
	\caption{Sparsity plots of the adjacency matrix for the friendship layer, with rows and columns ordered according to their community assignments in the consensus partitions for (a) group 1 and (b) group 2.}
	\label{fig:lazegaConsensusSpy}
\end{figure}

We use consensus clustering \cite{lancichinetti2012} with the null model from \cite{jeub2015thesis} to extract a representative partition for the trials in each of the two labeled groups, and we visualize these consensus partitions in Fig.~\ref{fig:lazegaConsensus}. 
As expected from Fig.~\ref{fig:lazegaScatterNMI}(a), the consensus partition for group $2$ has identical communities across the three layers. The structure of the consensus partition for group $1$ is more interesting, and it reveals a set of lawyers who are in a different community in the friendship layer than in the co-work and advice layers. In fact, the friendship layer is what distinguishes the two partitions from Fig.~\ref{fig:lazegaConsensus}(a) and Fig.~\ref{fig:lazegaConsensus}(b). In Fig.~\ref{fig:lazegaConsensusSpy}, we show sparsity plots of the adjacency matrix for the friendship layer; we order the rows and columns of this matrix according to the two consensus partitions from Fig.~\ref{fig:lazegaConsensus}. The three diagonal blocks in each panel correspond to the three communities. There is a larger number of intra-community entries in Fig.~\ref{fig:lazegaConsensusSpy}(a) than in Fig.~\ref{fig:lazegaConsensusSpy}(b). This suggests that the consensus partition for group 1 identifies stronger assortative structure in the friendship layer than the consensus partition for group 2. This simple example 
illustrates a key trade-off in multilayer community detection between detecting ``optimal" communities in each layer (as the partition for group 1 tends to do) and detecting communities that are stable across layers (as is the case for the consensus partition for group 2).

\begin{table}[ht!]
	\centering
	\caption{NMI scores between metadata and consensus partitions for the Lazega Law Firm data set. We highlight the largest value in each column.}
	\label{table:lazegaNMI}
	\begin{tabular}{l|ccccccc}
		\hline\hline
		& Office &Practice &Age &Seniority &Status &Gender &Law school \\ \hline
		Partition 1 & 0.587 & 0.334 & \highlight{0.146} & \highlight{0.147} & \highlight{0.150} & 0.035 & \highlight{0.024} \\ 
		Partition 2 & \highlight{0.610} & \highlight{0.469} & 0.098 & 0.052 & 0.040 & 0.026 & 0.007 \\ \hline
		All 100 runs & 0.577 & 0.406 & 0.125 & 0.106 & 0.093 & \highlight{0.037} & 0.022 \\ \hline\hline
	\end{tabular}
\end{table}

We can gain further insight into the two consensus partitions by computing NMI scores with each type of metadata. We report our results in Table \ref{table:lazegaNMI}. For comparison, the bottom row of this table gives the mean NMI score across all $100$ runs of our iterative algorithm. Our results are in line with a previous study of this data set \cite{peel2017ground}, which found that the office and practice metadata are more strongly related to network community structure than the law-school or gender metadata. We also observe that the two consensus partitions tend to correlate with different pieces of metadata. From Fig.~\ref{fig:lazegaConsensus}, we know that these differences are driven by variations in the friendship layer. This suggests that age, seniority, and status are more important determinants of friendship than of co-work and advice relationships. By contrast, office location and type of practice (litigation or corporate) are more significant drivers of co-work and advice relationships than of friendships.

As this example shows, there is no guarantee that our {\sc IterModMax} algorithm identifies a single set of optimal parameter values and thus a single multilayer assignment of a network's nodes to communities. However, we feel that this is a strength, rather than a shortcoming, of our method. Many empirical networks may have multiple meaningful partitions into communities, which, for example, may align best with different pieces of metadata \cite{peel2017ground}. Even without the guarantee of a unique output, our method gives a systematic and principled way to reduce the number of cases to investigate using a much smaller number of computations than other similar methods \cite{weir2017post}.


\section{Conclusions}\label{sec:conclusions}

The ability to coarse-grain a network by identifying its constituent communities is important for many applications. Many recent developments in community detection have focused on multilayer networks, which one can use to encode time-dependent interactions, multiple types of interactions, and other complications that can arise in complex systems. There exist a variety of methods for community detection in multilayer networks. In this paper, we focused on two approaches: modularity maximization and statistical inference in stochastic block models. Our main results connect these two types of methods. By considering various types of multilayer structure, we showed that the multilayer modularity objective function is, under specific circumstances, the same as the posterior log-probability that corresponds to certain multilayer SBMs. Therefore, maximizing one expression with respect to community assignments is equivalent to maximizing the other. This link between the two approaches highlights some implicit assumptions of classical multilayer modularity that previously were not clear. These include the assumption that layers are ``statistically equivalent" (which means that they are described by the same sets of parameters, $\theta_\mathrm{in}$ and $\theta_\mathrm{out}$) and that all communities in a layer have the same expected sizes (because the null distributions $\Pp_0^t$ are uniform and the copying probabilities $p_t$ are the same for all nodes in a layer).

For temporal networks, we examined both the situation in which the resolution and interlayer-coupling parameters are uniform across all layers and the one in which they are layer-dependent. In the latter case, we proposed a novel layer-weighted version of modularity that is appropriate for situations in which communities in different layers have different statistical properties. We also briefly explored the use of layer-dependent models to detect structural change points \cite{peel2015detecting}, which correspond to layers in which a network's mesoscale structure undergoes a major reorganization (see Sec. \ref{sm:sec:numericalExamples} of our supplementary materials). For multiplex networks, we proposed a new way of generating multilayer partitions by permuting the layers and then updating community labels in the order indicated by the permutation. In contrast to other generative models of multiplex networks \cite{bazzi2016generative}, our approach makes it possible to write down the probability of generating a particular multilayer partition in closed form, and we expect that this will be useful for future efforts (e.g., for performing inference in the resulting SBMs).

Our work also suggests a principled method for determining appropriate values for the resolution and interlayer-coupling parameters that arise in the formulation of multilayer modularity. 
Our iterative algorithm provides a way to explore parameter space in a more efficient way than an exhaustive grid search, and we can thereby use fewer computations than previous approaches \cite{bassett2013,weir2017post}. 
We tested our algorithm on two families of multilayer benchmark networks. Our results showed the expected qualitative features, with better performance in networks with stronger community structure and more persistent communities across layers. We also used our algorithm on the Lazega Law Firm data set, which is a multiplex network with three layers. Our analysis revealed a set of lawyers who belong to a community in the friendship layer that is different from their community in the co-work and advice layers. 

There are many possible extensions of our work. One can derive similar results for multilayer networks with different forms of interlayer coupling from those that we discussed in this paper. Examples include temporal networks with memory, in which a layer depends not just on its predecessor, but on all other layers that came before it; networks with community-dependent coupling parameters; interconnected networks, in which interlayer connections are no longer diagonal; and multilayer networks with more than one aspect (e.g., both temporal and multiplex ones \cite{kivela2014}). More broadly, our work (as well as Newman's previous work \cite{newman2016}) provides a recipe for deriving new versions of modularity for both monolayer and multilayer networks by starting from an SBM and making appropriate simplifications. Previous authors have used such a procedure to propose a version of modularity that incorporates information from metadata \cite{xie2017} by simplifying the SBM from \cite{newman2016annotated}. We expect that one can similarly derive modularity functions from other SBMs that incorporate node annotations \cite{hric2016metadata,zhang2016nodefeatures} in both monolayer and multilayer settings.
Another idea is to incorporate parameters that control expected community sizes in the definition of modularity, effectively relaxing the assumption that the null distributions $\Pp_0^t$ are uniform \cite{zhang2016community}.

We also anticipate that it is possible to improve on Alg.~\ref{alg:itModMax}. For example, rather than using a single run of modularity maximization to estimate new resolution and interlayer-coupling parameter values, one can combine information from multiple runs to choose a direction to move and a step size to use in a move in that direction. Another challenging task is to extend the results of the present paper to weighted networks. A key observation is that such weights only impact intralayer terms. Therefore, to study extensions of our analysis to weighted networks, it is sufficient to relate weighted modularity to a suitable monolayer SBM for weighted networks \cite{aicher2014,peixoto2018}. 

In summary, we showed that two common and independently-developed methods for community detection in multilayer networks --- modularity maximization and SBM-based statistical inference --- are closely related. This connection goes far beyond being only an interesting theoretical finding, as we believe that the techniques that we developed in the present paper can provide practical insights into devising more robust community-detection algorithms that do not require ad-hoc experimentation with parameter values.

\begin{acknowledgments}

ARP was funded by the EPSRC Centre for Doctoral Training in Industrially Focused Mathematical Modelling (EP/L015803/1) in partnership with dunnhumby. We thank Marya Bazzi, Lucas Jeub, and Florian Klimm for useful discussions.

\end{acknowledgments}


\appendix

\section{The Relative Weight of Intralayer and Interlayer Contributions for Undirected and Directed Networks}\label{sm:sec:directed}

In this section, we compare the optimal modularity functions that one obtains for undirected and directed networks using our approach from Sec.~\ref{sec:multilayerEquiv}. These two objective functions should be identical when the intralayer adjacency matrices $\mat{A}^t$ are symmetric, as treating each undirected edge as two directed edges should not change the objective function that one optimizes for community detection. To simplify our discussion, we focus on the uniform setting (in which all layers are described by the same parameters).

Recall from Sec.~\ref{subsec:multilayerSBMs} that the log-likelihood for undirected multilayer networks under the PPM is
\begin{align}\label{eqn:sm:likelihoodUndir}
	\log\Pp(\matt{A}|\vec{g},\theta_\mathrm{in},\theta_\mathrm{out}) 
	&=\left(\log\theta_\mathrm{in}-\log\theta_\mathrm{out}\right)\sum_{t=1}^T{\sum_{i,j=1}^N{
			\left( A_{ij}^t-\frac{\theta_\mathrm{in}-\theta_\mathrm{out}}{\log\theta_\mathrm{in}-\log\theta_\mathrm{out}}\frac{d_i^td_j^t}{2m_t}\right)\delta(g_i^t,g_j^t)}} \notag \\
	&\qquad+(\mbox{const.})  \,.
\end{align}
As in Eqn. \eqref{eqn:multilayerLikelihoodUnif}, the fact that this expression involves a sum over all $i$ and $j$, and not just over $i \leq j$, entails that the entries $A_{ij}^t$ and $A_{ji}^t$ of an intralayer adjacency matrix both contribute to the log-likelihood. A similar derivation for directed networks gives
\begin{align}\label{eqn:sm:likelihoodDir}
	\log\Pp(\matt{A}|\vec{g},\theta_\mathrm{in},\theta_\mathrm{out}) 
	&=\left(\log\theta_\mathrm{in}-\log\theta_\mathrm{out}\right)\sum_{t=1}^T{\sum_{i,j=1}^N{
			\left( A_{ij}^t-\frac{\theta_\mathrm{in}-\theta_\mathrm{out}}{\log\theta_\mathrm{in}-\log\theta_\mathrm{out}}\frac{d_{i,\mathrm{out}}^td_{j,\mathrm{in}}^t}{m'_t}\right)\delta(g_i^t,g_j^t)}} \notag \\
	&\qquad+(\mbox{const.})  \,,
\end{align}
where $d_{i,\mathrm{out}}^t$ is the out-degree of node $i$ in layer $t$, the quantity $d_{j,\mathrm{in}}^t$ is the in-degree of node $j$ in layer $t$, and $m'_t=\sum_{i,j=1}^N{A_{ij}^t}$ is the number of directed edges in layer $t$. When using a PPM to generate directed networks, the number of directed edges from node $i$ to $j$ in layer $t$ follows a Poisson distribution whose mean is either $d_{i,\mathrm{out}}^td_{j,\mathrm{in}}^t\theta_\mathrm{in}/m'_t$ or $d_{i,\mathrm{out}}^td_{j,\mathrm{in}}^t\theta_\mathrm{out}/m'_t$, depending on whether or not $i$ and $j$ are in the same community in layer $t$. The expression \eqref{eqn:sm:likelihoodDir} corresponds to a sum of monolayer modularities with the directed null model from \cite{leicht2008}. 

If the adjacency matrices $\mat{A}^t$ are all symmetric, $d_{i,\mathrm{out}}^t=d_{i,\mathrm{in}}^t=d_i^t$, $d_{j,\mathrm{out}}^t=d_{j,\mathrm{in}}^t = d_j^t$, and $m'_t=\sum_{i,j=1}^N{A_{ij}^t}=2m_t$. (The number of directed edges is twice the number of undirected edges.) The expressions \eqref{eqn:sm:likelihoodUndir} and \eqref{eqn:sm:likelihoodDir} become identical. When adding the log-prior term $\log\Pp(\vec{g}|p,K)$ and comparing the resulting expression with multilayer modularity, one then obtains the same optimal value of $\omega$ in both the undirected and directed settings. The assumption that $A_{ij}^t$ and $A_{ji}^t$ both contribute to the log-likelihood \eqref{eqn:sm:likelihoodUndir} for undirected networks is a crucial one: if one were to introduce a factor of $1/2$ to discount this double-counting of terms, the resulting optimal value of $\omega$ would be twice as large when assuming that the matrices $\mat{A}^t$ encode undirected networks rather than directed ones.

It is important to note that one of our paper's key results, which relates the intralayer component of modularity to a sum of log-likelihoods and the interlayer component to a log-prior on community assignments, holds irrespective of the relative scaling between these two types of contributions. It is only when one uses this relation for parameter estimation (see Sec.~\ref{sec:parameterEstimation}) that the formula for $\omega$ depends on this scaling. For our numerical experiments from Sec.~\ref{subsec:numericalExamples} and Sec.~\ref{sm:sec:numericalExamples}, we have found that performance deteriorates if one rescales the log-likelihood \eqref{eqn:sm:likelihoodUndir} by a factor of $1/2$ to avoid double-counting edges. It is possible that one would see improved performance in our parameter-estimation algorithm by incorporating an additional tuning constant $\alpha$ and maximizing the modularity function that is equivalent (for suitable $\gamma$ and $\omega$) to the expression 
\begin{equation*}
	\log\Pp(\matt{A}|\vec{g},\theta_\mathrm{in},\theta_\mathrm{out}) + \alpha\log\Pp(\vec{g}|p,K)\,.
\end{equation*} 
Such a constant is analogous to ones that have been introduced for community detection in the presence of metadata \cite{emmons2018,murphy2016}. It is also akin to regularization constants that appear in other types of optimization problems, as the term $\log\Pp(\vec{g}|p,K)$ helps prevent overfitting to the community structure of individual layers. Investigating the effect of such a trade-off parameter is beyond the scope of the present paper, but it is an interesting topic for future work.


\section{Derivation of Theoretically-Optimal Parameter Values for Multilevel Networks}\label{sm:sec:multilevel}

In a multilevel network \cite{lomi2016}, the layers are a hierarchy of monolayer networks, and interlayer edges between nodes indicate inclusion relationships. Like the temporal networks from Sec.~\ref{subsec:temporal} of the main manuscript, the layers in multilevel networks have a natural ordering, which makes it straightforward to modify the approach from Sec.~\ref{subsec:temporal} and obtain a similar result. This section complements our results for temporal and multiplex networks from Sec.~\ref{subsec:temporal} and Sec.~\ref{subsec:multiplex}, respectively.

Multilevel networks have been studied before, most notably by Snijders and collaborators, who used multilevel modeling for social-network analysis \cite{snijders2003multilevel,snijders1995,van1999multilevel}. The use of multilevel networks for community-detection applications is considerably rarer. We note a recent study by Barbillon et al. \cite{barbillon2017}, who investigated a data set of collaborations and exchanges of resources between cancer researchers in France and their laboratories. Although Barbillon et al. modeled these interactions as a multiplex network of researchers, one can alternatively use a multilevel representation with researchers in one layer and laboratories in the other layer. A related example is a multilayer collaboration network, in which coauthored papers induce connections between individuals, research groups, departments, and universities (or a subset of these). A third potential application is international trade networks, in which nodes at higher levels represent larger geographical areas. 

A key difference between multilevel networks and the other examples that we consider in this paper (including temporal and multiplex networks) is that we need to consider non-diagonal interlayer edges. Additionally, nodes in different layers no longer necessarily correspond to the same set of entities. We therefore need to introduce some new notation. Let $\mathcal{N}^t$ be the set of nodes in layer $t$, which has $|\mathcal{N}^t|=N^t$ nodes. Assume that $\mathcal{N}^1$ corresponds to the top level of the hierarchy. For $t \in \{2,\ldots,T\}$, the functions $\pi^t:\mathcal{N}^t \rightarrow \mathcal{N}^{t-1}$ send nodes $i \in \mathcal{N}^t$ to their parents immediately higher up in the hierarchy. (For example, such a function can map researchers to their departments.) For convenience, we use the notation $\pi_i^t$ to denote $\pi^t(i)$; note the implicit assumption that node $i$ is from layer $t$.

The log-likelihood for this multilevel network,
\begin{align}\label{eqn:multilevelLikelihoodNonUnif}
	&\log\Pp(\matt{A}|\vec{g},\vec{\theta}_\mathrm{in},\vec{\theta}_\mathrm{out})  
	= \sum_{t=1}^T\Biggl[\left(\log\theta_\mathrm{in}^t-\log\theta_\mathrm{out}^t\right)
	\sum_{i,j \in \mathcal{N}^t}{\left( A_{ij}^t-\frac{\theta_\mathrm{in}^t-\theta_\mathrm{out}^t}{\log\theta_\mathrm{in}^t-\log\theta_\mathrm{out}^t}\frac{d_i^td_j^t}{2m_t}\right)\delta(g_i^t,g_j^t)}\Biggr] \\
	&\qquad +\mbox{(const.)}\,, \nonumber
\end{align}
is similar to the one in Eqn.~\eqref{eqn:temporalLikelihoodNonUnif} for non-uniform temporal networks. The prior on $\vec{g}$ is also similar to that for temporal networks, although information flows from the top level of the hierarchy to the bottom level, rather than in the direction of time. First, we generate the vector $\vec{g}^1$ of community assignments in the top level by sampling from the null distribution $\Pp_0^1$. To then sample community assignments $\vec{g}^t$ in subsequent layers, we 
assume that each node copies its community label from its parent with probability $p_t$ and otherwise randomly samples a label from a null distribution $\Pp_0^t$. This process yields the following log-prior:
\begin{align}\label{eqn:multilevelLogPrior}
	\log\Pp(\vec{g}) 
	&= \sum_{i \in \mathcal{N}^1}{\log\Pp_0^1(g_i^1)}+\sum_{t=2}^T\sum_{i\in \mathcal{N}^t}{\log\left[(1-p_t)\Pp_0^t(g_i^t) \right]} \notag \\
	&\qquad+\sum_{t=2}^T\sum_{i\in \mathcal{N}^t}{\log\left[ 1+\frac{p_t}{(1-p_t)\Pp_0^t(g_i^t)} \right]\delta(g_{\pi_i^t}^{t-1},g_i^t)}\,. 
\end{align}
As in Sec.~\ref{subsec:temporal}, we assume that the null distributions $\Pp_0^t$ are uniform, so the first two terms in the right-hand side of \eqref{eqn:multilevelLogPrior} are independent of $\vec{g}$. In particular, we let $\Pp_0^t(g_i^t)=1/K_t$. Ignoring constants, we then obtain the following expression for the log-prior:
\begin{align}\label{eqn:priorMultilevel}
	\log\Pp(\vec{g})=\sum_{t=2}^T\sum_{i\in \mathcal{N}^t}{\log\left( 1+\frac{p_t}{1-p_t}K_t \right)\delta(g_{\pi_i^t}^{t-1},g_i^t)} + \mbox{(const.)} \,.
\end{align}
From Eqns.~\eqref{eqn:multilevelLikelihoodNonUnif} and \eqref{eqn:priorMultilevel}, it follows that the posterior distribution for $\vec{g}$ is
\begin{align}\label{eqn:posteriorMultilevel}
	&\log\Pp(\vec{g}|\matt{A},\vec{\theta}_\mathrm{in},\vec{\theta}_\mathrm{out},\vec{p},\vec{K}) \notag \\
	&\qquad=\sum_{t=1}^T\Biggl[\left(\log\theta_\mathrm{in}^t-\log\theta_\mathrm{out}^t\right) 
	\sum_{i,j \in \mathcal{N}^t}{\left( A_{ij}^t-\frac{\theta_\mathrm{in}^t-\theta_\mathrm{out}^t}{\log\theta_\mathrm{in}^t-\log\theta_\mathrm{out}^t}\frac{d_i^td_j^t}{2m_t}\right)\delta(g_i^t,g_j^t)}\Biggr] \notag \\
	&\qquad\qquad + \sum_{t=2}^T\sum_{i\in \mathcal{N}^t}{\log\left( 1+\frac{p_t}{1-p_t}K_t \right)\delta(g_{\pi_i^t}^{t-1},g_i^t)} + \mbox{(const.)} \,.
\end{align}

The corresponding layer-weighted modularity for multilevel networks is
\begin{equation}\label{eqn:modularityMultilevel}
	Q(\vec{g}) = \sum_{t=1}^T{\beta_t\sum_{i,j \in \mathcal{N}^t}{\left(A_{ij}^t-\gamma_t\frac{d_i^td_j^t}{2m_t}\right)\delta(g_i^t,g_j^t)}} +\sum_{t=2}^T\sum_{i \in \mathcal{N}^t}{\omega_t\delta(g_{\pi_i^t}^{t-1},g_i^t)}\,.
\end{equation}

Comparing the expressions in \eqref{eqn:posteriorMultilevel} and \eqref{eqn:modularityMultilevel}, we arrive at the following values for the resolution parameters $\gamma_t$, coupling parameters $\omega_t$, and layer weightings $\beta_t$:
\begin{align}
	\gamma_t &= \frac{\theta_\mathrm{in}^t-\theta_\mathrm{out}^t}{\log\theta_\mathrm{in}^t-\log\theta_\mathrm{out}^t} \,, \\
	\omega_t &= \frac{1}{\langle \log\theta_\mathrm{in}^t-\log\theta_\mathrm{out}^t \rangle_t}\log\left( 1+\frac{p_t}{1-p_t}K_t \right)\,, \\
	\beta_t &= \frac{\log\theta_\mathrm{in}^t-\log\theta_\mathrm{out}^t}{\langle \log\theta_\mathrm{in}^t-\log\theta_\mathrm{out}^t \rangle_t} \,.
\end{align}
These are the same expressions as the ones for temporal networks in Eqns.~\eqref{eqn:gammaOptimalNonUnif}--\eqref{eqn:betaOptimalNonUnif}. Additionally, just as in the temporal case, one can also analyze the case of layer-independent parameters $\gamma$ and $\omega$. (In that case, the layer weights $\beta_t$ are equal to $1$.) Although this derivation is very similar to the one for temporal networks from Sec.~\ref{subsec:temporal}, the application to multilevel networks is markedly different, due to the interpretation of interlayer edges as inclusion relationships. Community detection in multilevel networks has been studied sparsely thus far, and we hope to see more such applications in the future.

\section{Estimating SBM Parameters}\label{sm:sec:estimateSBM}

In this section, we explain how to estimate, given a multilayer partition $\vec{g}$, the parameters of our various SBMs from Sec.~\ref{sec:multilayerEquiv}. This completes the description of our iterative modularity-maximization algorithm from Sec.~\ref{subsec:iterativeAlg.}. 

We approximate $\theta_\mathrm{in}$ and $\theta_\mathrm{out}$ in a manner similar to that presented in \cite{newman2016} for monolayer networks. Let $M_\mathrm{in}^t$ denote the random variable for the number of intra-community edges in layer $t$ under the degree-corrected SBM from Sec.~\ref{subsec:multilayerSBMs}. The expected number of such edges across all layers is
\begin{equation}\label{eqn:expectedMIn}
	\E\left[\sum_{t=1}^T{M_\mathrm{in}^t}\right]=\frac{1}{2}\sum_{t=1}^T{\sum_{i,j=1}^N{\theta_\mathrm{in}\frac{d_i^t d_j^t}{2m_t}\delta(g_i^t,g_j^t)}}  
	= \frac{1}{2}\theta_\mathrm{in}\sum_{t=1}^T{\frac{\sum_{r}{\left(\kappa_{r}^t\right)^2}}{2m_t}}\,,
\end{equation}
where $\kappa_r^t$ denotes the sum of the degrees of nodes in community $r$ in layer $t$. Replacing the expectation on the left-hand side of Eqn.~\eqref{eqn:expectedMIn} with the observed number of intra-community edges ($\sum_{t=1}^T{m_\mathrm{in}^t}$) yields the estimate
\begin{equation}\label{eqn:thetaInEstimate}
	\theta_\mathrm{in} \approx \frac{\sum_{t=1}^T{2m_\mathrm{in}^t}}{\sum_{t=1}^T{\frac{1}{2m_t}\sum_{r}{\left(\kappa_{r}^t\right)^2}}}\,.
\end{equation}
Using a similar approach, we estimate that
\begin{equation}\label{eqn:thetaOutEstimate}
	\theta_\mathrm{out} \approx \frac{\sum_{t=1}^T{2m_\mathrm{out}^t}}{\sum_{t=1}^T{\left[2m_t-\frac{1}{2m_t}\sum_{r}{\left(\kappa_{r}^t\right)^2}\right]}}\,.
\end{equation}
In the layer-dependent case, one can similarly determine each set of parameters, $\theta_\mathrm{in}^t$ and $\theta_\mathrm{out}^t$, by considering intra-community and inter-community edges separately for each layer.

We now consider the other two sets of parameters. We set the number $K$ of communities equal to the observed number of distinct community labels in the most recent multilayer partition $\vec{g}$. That is,
\begin{equation}\label{eqn:KEstimate}
	K=\left|\bigl\{g_i^t:i \in \{1,\ldots,N\}\,, t\in \{1,\ldots,T\}\bigr\}\right|\,.
\end{equation}
In the layer-dependent case, we set
\begin{equation}\label{eqn:KEstimateNonunif}
	K_t=\left|\bigl\{g_i^t:i \in \{1,\ldots,N\}\bigr\}\right|\,, \quad t\in \{1,\ldots,T\}\,.
\end{equation}
Estimating the copying probabilities $p$ (or $p_t$, if they are layer-dependent) is more difficult. The details of the calculation also depend on the type of multilayer network that one considers. We discuss temporal and multilevel networks in Sec.~\ref{subsubsec:temporalMultilevelEstimation} and multiplex networks in Sec.~\ref{subsubsec:multiplexEstimation}. 


\subsection{Temporal and multilevel networks}\label{subsubsec:temporalMultilevelEstimation}

To estimate the copying probability $p$, note that the probability that a node $i$ is in the same community in two consecutive layers in our model is 
\begin{equation}\label{eqn:persModel}
	\Pp(g_i^{t-1}=g_i^t)=p+\frac{1-p}{K}\,.
\end{equation}
The first term corresponds to the probability that node $i$ copies its community assignment from the previous layer, and the second term is the probability that node $i$ samples that community uniformly at random from the $K$ available choices. Estimating the same probability empirically from network data amounts to counting the number of times that a node stays in the same community (i.e., ``persists") across two consecutive layers:
\begin{equation}\label{eqn:persData}
	\Pp(g_i^{t-1}=g_i^{t}) \approx \frac{1}{N(T-1)}\sum_{t'=2}^T{\sum_{i'=1}^N{\delta(g_{i'}^{t'-1},g_{i'}^{t'})}}=\frac{\Pers(\vec{g})}{N(T-1)}\,,
\end{equation}
where we follow the notation in \cite{bazzi2016} and let $\Pers(\vec{g})$ denote the number of instances in which a node belongs to the same community in consecutive layers. Equating the empirical probability \eqref{eqn:persData} to the theoretical one in \eqref{eqn:persModel} yields
\begin{equation}\label{eqn:pEstimate}
	p+\frac{1-p}{K} \approx \frac{\Pers(\vec{g})}{N(T-1)} \quad\Rightarrow\quad p \approx \left[\frac{\Pers(\vec{g})}{N(T-1)}-\frac{1}{K}\right] \bigg/ \left(1-\frac{1}{K}\right)\,.
\end{equation}

In the layer-dependent case, a similar derivation yields
\begin{equation}
	p_t \approx \left[\frac{1}{N}\sum_{i=1}^N{\delta(g_i^{t-1},g_i^t)}-\frac{1}{K}\right] \bigg/ \left(1-\frac{1}{K}\right)\,, \quad t \in \{2,\ldots,T\}\,.
\end{equation}

We follow an almost identical approach for multilevel networks. In the uniform case, our estimate for $p$ is the same as in Eqn.~\eqref{eqn:pEstimate}, although the calculation of persistence $\mbox{Pers}(\vec{g})$ is different:
\begin{equation}
	\mbox{Pers}(\vec{g})=\sum_{t=2}^T\sum_{i \in \mathcal{N}^t}{\delta(g_{\pi_i^t}^{t-1},g_i^t)} \quad\Rightarrow\quad p \approx \left[\frac{1}{T-1}\sum_{t=2}^T\frac{1}{N^t}\sum_{i \in \mathcal{N}^t}{\delta(g_{\pi_i^t}^{t-1},g_i^t)}-\frac{1}{K}\right] \bigg/ \left(1-\frac{1}{K}\right)\,.
\end{equation}
Similarly, for the layer-dependent case, 
\begin{equation}
	p_t \approx \left[\frac{1}{N^t}\sum_{i \in \mathcal{N}^t}{\delta(g_{\pi_i^t}^{t-1},g_i^t)}-\frac{1}{K}\right] \bigg/ \left(1-\frac{1}{K}\right)\,, \quad t \in \{2,\ldots,T\}\,.
\end{equation}


\subsection{Multiplex networks}\label{subsubsec:multiplexEstimation}

Estimating the probability $p$ for the multiplex version of our SBM --- or the probabilities $p_{st}$, if one is considering layer-dependent parameters --- is significantly more difficult than for temporal or multilevel networks. We start by considering the uniform case.

Consider the probability $\Pp(g_i^s=g_i^t)$ for a randomly chosen node $i$ and two arbitrary layers $s$ and $t$ (with $s \neq t$). For a given multilayer partition $\vec{g}$, we estimate $\Pp(g_i^s=g_i^t)$ by counting the number of instances in which two community labels are the same and dividing by the total number of instances:
\begin{equation}\label{eqn:multiplexpEmpirical}
	\Pp(g_i^s=g_i^t) \approx \frac{1}{NT(T-1)}\sum_{i'=1}^N\sum_{t'=1}^T\sum_{s' \neq t'}{\delta(g_{i'}^{s'},g_{i'}^{t'})}\,.
\end{equation}

\begin{figure}[h!]
	\centering
	\medskip
	\includegraphics[width=0.48\textwidth]{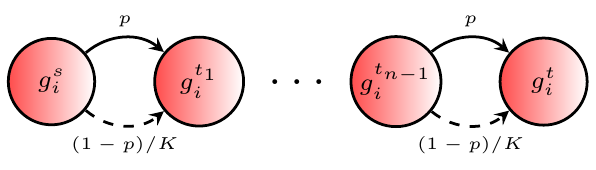}
	\caption{Illustration of the copying and sampling process that propagates community labels across layers. Assume that after applying a permutation, the order of the layers is $s,t_1,\ldots,t_{n-1},t$. We are interested in the probability that a node $i$ has the same community labels in layers $s$ and $t$ (i.e., that $g_i^s=g_i^t$). For each layer from $t_1,\ldots,t_{n-1},t$, node $i$ copies its community label from the previous layer with probability $p$ (solid arrows), or it samples the same label uniformly at random with probability $(1-p)/K$ (dashed arrows).}
	\label{fig:multiplexDiagram}
\end{figure}

We now calculate $\Pp(g_i^s=g_i^t)$ for the generative model that we described in Sec. \ref{subsec:multiplex}. To do this, we need to account for all of the possible ways in which a node $i$ can have the same community label in layers $s$ and $t$. Suppose that we update labels in the order that is indicated by a permutation $\sigma$ and that, in this permutation, there are $n-1$ intermediate layers between $s$ and $t \equiv t_n$. We denote these layers by $t_1,\ldots,t_{n-1}$. By symmetry, we may assume that layer $s$ comes before layer $t$ in the permutation, and we introduce a factor of $2$ when necessary to account for the reverse case. 
We can break down the probability $\Pp(g_i^s=g_i^t)$ into two components: either (1) node $i$ in layer $t$ copies the label from layer $t_{n-1}$ and that label is identical to the one from layer $s$; or (2) node $i$ in layer $t$ randomly samples the same label from layer $s$. Therefore, we obtain the recursive relationship
\begin{equation}\label{eqn:multiplexRecursion1}
	\Pp(g_i^s=g_i^{t_n})=p\Pp(g_i^s=g_i^{t_{n-1}})+\frac{1-p}{K}\,,
\end{equation} 
which holds for any $n$. We illustrate this process in Fig.~\ref{fig:multiplexDiagram}. The base case is
\begin{equation}\label{eqn:multiplexRecursion2}
	\Pp(g_i^s=g_i^{t_1})=p+\frac{1-p}{K} \,.
\end{equation}

Let $f(x)=px+(1-p)/K$. Thus, $f(1)$ is the probability that a node in one layer either copies the community label from the previous layer (after applying the permutation) or that it randomly samples the same label. We can then summarize the recursion from Eqns.~\eqref{eqn:multiplexRecursion1} and \eqref{eqn:multiplexRecursion2} with
\begin{equation}\label{this_expression}
	\Pp(g_i^s=g_i^{t_n})=f^{(n)}(1)\,,
\end{equation}
where $f^{(n)}$ denotes $f$ composed with itself $n$ times.

The expression \eqref{this_expression} corresponds to the probability that nodes $s$ and $t$ share the same label when they are $n$ layers apart after applying a permutation $\sigma$. To get the overall probability, we first have to count, for each $n$, the number of permutations with this property. Specifically, we need to consider all $\sigma \in S_T$ with $\sigma(u)=s$, $\sigma(u+1)=t_1$, $\ldots,$ $\sigma(u+n)=t_n \equiv t$ for some $u$. Using straightforward combinatorics, one can show that there are $(T-2)!(T-n)$ such permutations. This yields
\begin{equation}\label{above}
	\Pp(g_i^s=g_i^t)= \sum_{n=1}^{T-1}{\frac{2(T-2)!(T-n)}{T!}f^{(n)}(1)} 
	= \sum_{n=1}^{T-1}{\frac{2(T-n)}{T(T-1)}f^{(n)}(1)} \,.
\end{equation}

For $p=1$, note that $f(1)=1$, so
Eqn.~\eqref{above} gives $\Pp(g_i^s=g_i^t)=1$, as expected. For $p<1$, the function $f$ has a unique fixed point at $1/K$. We can write 
\begin{equation*}
	f^{(n)}(1)-\frac{1}{K}=p\left( f^{(n-1)}(1)-\frac{1}{K}\right)=\cdots=p^n\left( 1-\frac{1}{K} \right)\,,
\end{equation*}
which implies that
\begin{equation*}
	f^{(n)}(1)=p^n\frac{K-1}{K}+\frac{1}{K}\,.
\end{equation*}
This yields
\begin{equation}\label{eqn:multiplexpModel}
	\Pp(g_i^s=g_i^t)=\frac{2(1-\frac{1}{K})}{T(T-1)}\sum_{n=1}^{T-1}{p^n(T-n)}+\frac{1}{K}\,.
\end{equation}
We can therefore estimate $p$ numerically by setting the right-hand side of Eqn.~\eqref{eqn:multiplexpEmpirical} to be equal to that of Eqn.~\eqref{eqn:multiplexpModel}. This leads to a polynomial root-finding problem that is easy to solve using readily available algorithms. It is also possible to rewrite \eqref{eqn:multiplexpModel} using formulas for the geometric series and its derivatives, but that complicates the numerics without giving additional insights into the estimated value of $p$. 

The above calculation does not work in the layer-dependent case. Previously, all permutations that put $n-1$ layers between $s$ and $t$ gave the same value for the probability $\Pp(g_i^s=g_i^t)$, but this is no longer the case when there are different copying probabilities $p_{st}$ for different layers. Instead, one needs to consider every permutation individually to obtain an exact expression for $\Pp(g_i^s=g_i^t)$. Such a brute-force approach is feasible for a small number of layers (say, $T \lessapprox 6$), but it will not work in general. One avenue for future research is exploring suitable approximations for estimating the parameters $p_{st}$ in this more complicated setting.


\section{Additional Numerical Examples}\label{sm:sec:numericalExamples}

Our examples in Sec.~\ref{subsec:numericalExamples} in the main manuscript consist of temporal networks in which each node has the same expected degree and each community has the same expected size. To test our methodology on networks with heterogeneous degree distributions and heterogeneous community sizes, both of which are common features of empirical networks, we turn to the general multilayer model in \cite{bazzi2016generative}. We use the code provided in \cite{benchmark} to generate networks using this generative model.

\begin{figure*}[htbp]
	\centering
	\subfloat[{\sc GenLouvain}]{\includegraphics[width=0.45\textwidth]{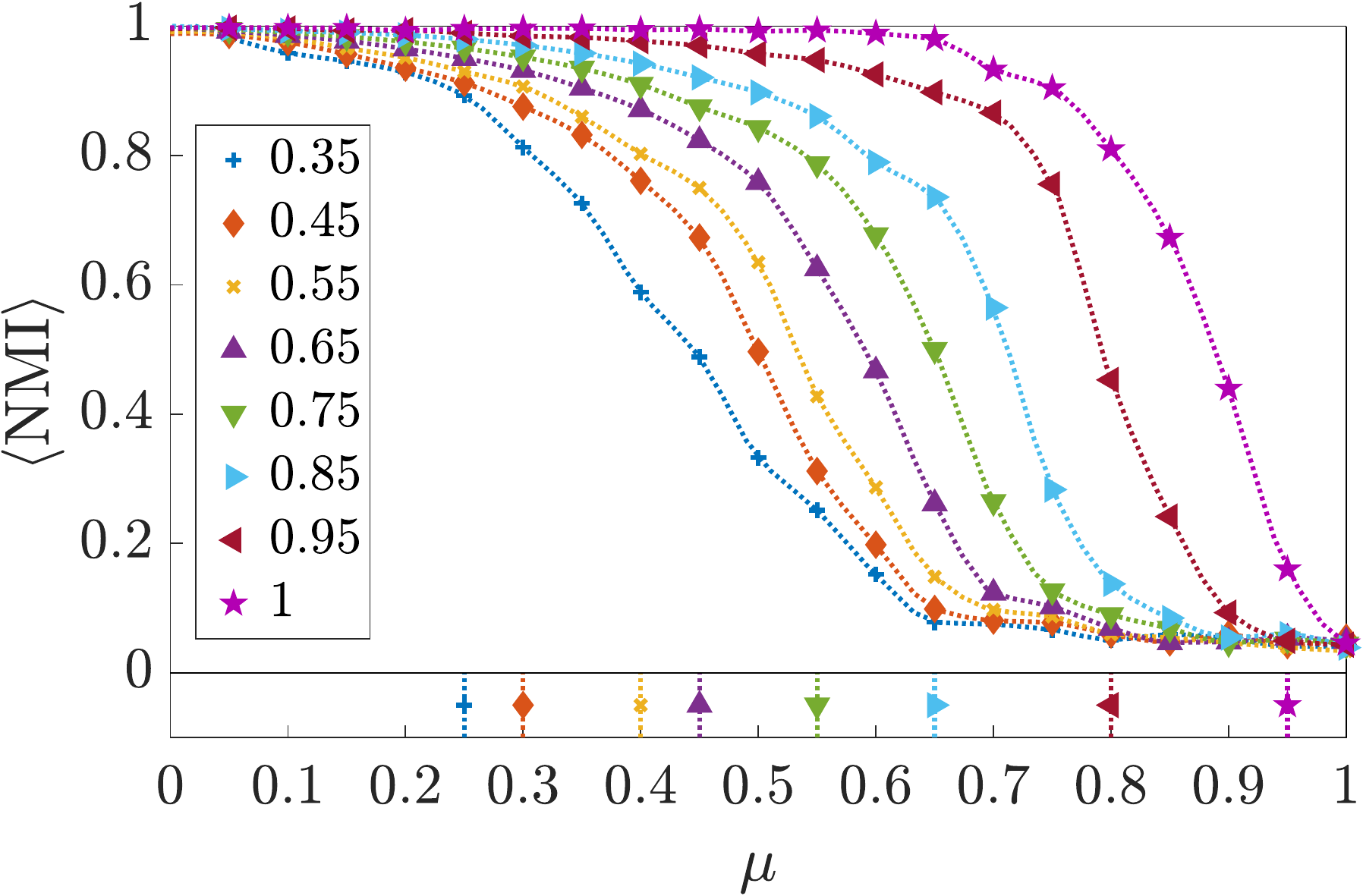}} \hfill
	\subfloat[{\sc GenLouvainRand}]{\includegraphics[width=0.45\textwidth]{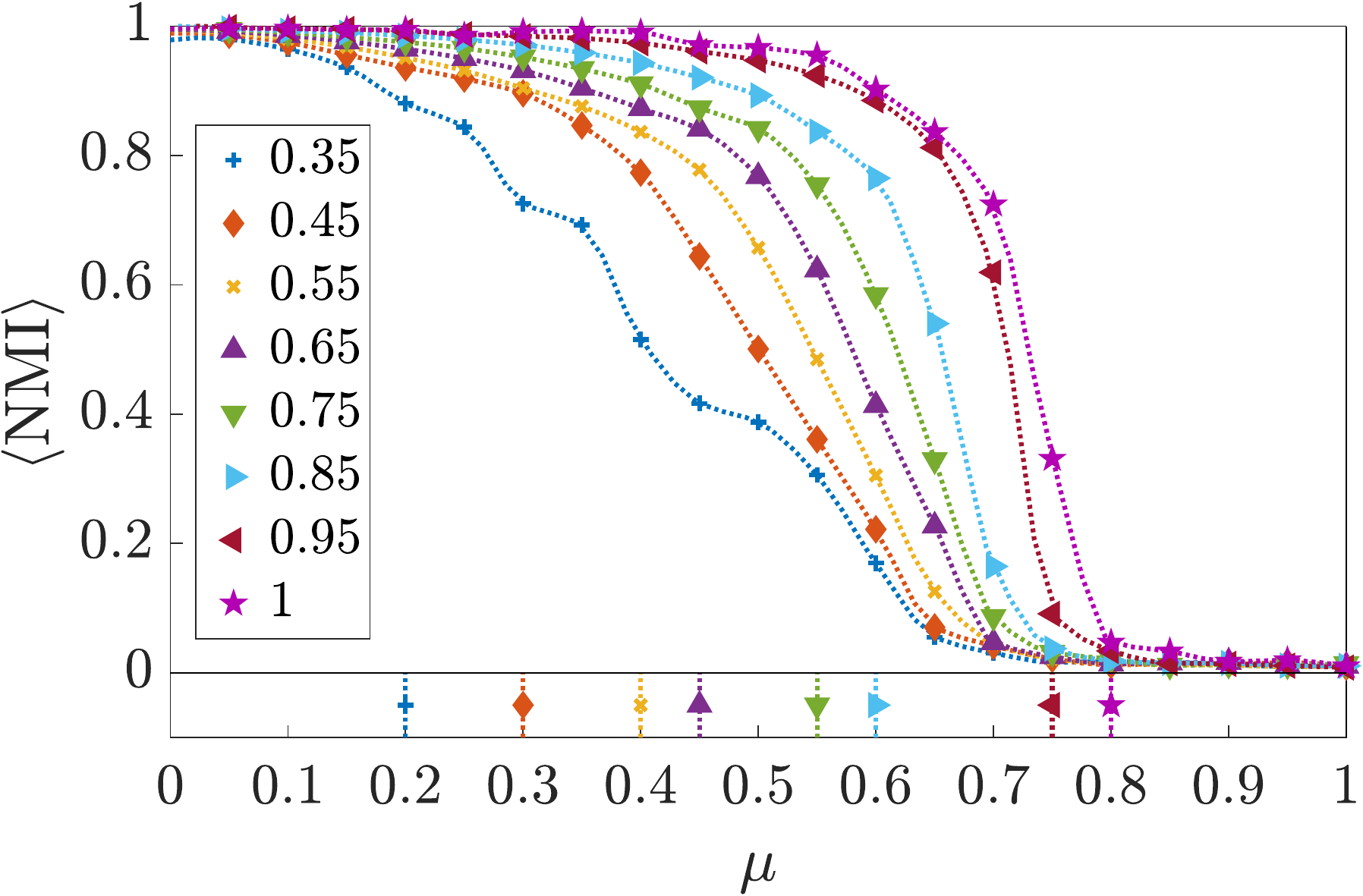}}
	\caption{Results for the temporal multilayer benchmark network from Bazzi et al. \cite{bazzi2016generative} using {\sc GenLouvain} and {\sc GenLouvainRand} to perform modularity maximization. We use the same parameter values that the authors used in their paper. The resulting networks have $T=100$ layers and $N=150$ nodes in each layer. There are $K=5$ communities, whose expected sizes follow a symmetric Dirichlet distribution with a concentration parameter with value $1$;
		and node degrees in each layer follow a truncated power-law distribution. (See \cite{bazzi2016generative} for details.) 
		Our plots show layer-averaged NMI scores between the planted partition and the one that we detect by modularity maximization as a function of the mixing parameter $\mu$. Each line and set of markers corresponds to a different value of the copying probability $p_b$, and each data point is a mean over $100$ trials. We use Alg.~\ref{alg:itModMax} to update the values of $\gamma$ and $\omega$ from the initial values
		$\gamma^{(0)}=1$ and $\omega^{(0)}=1$. The lines at the bottom of each plot indicate the values of $\mu$ up to which at least $10\%$ of the $100$ runs converge to a fixed point.}
	\label{fig:benchmarkTemporal}
\end{figure*}

Similar to the model of Ghasemian et al. \cite{ghasemian2016}, the simplest models from \cite{bazzi2016generative} have two parameters. A mixing parameter $\mu$ controls the strength of the planted community structure; when $\mu=0$, all edges lie within communities; when $\mu=1$, edges are equally likely to lie within communities and between communities. The second parameter of the benchmark from Bazzi et al. is the probability with which a node in one layer copies its community label from a different layer; we denote this probability by $p_b$ to distinguish it from our own parameter $p$. 
The parameters $p$ and $p_b$ are the same for temporal networks, but they are different for multiplex networks. The discrepancy in the latter case arises because our generative model from Sec.~\ref{subsec:multiplex} is different from the one from \cite{bazzi2016generative}. For multiplex networks, Bazzi et al. also introduced the quantity $\widehat{p}_b=(T-1)p_b$ for the probability that a node copies its community label from one of the other $T-1$ layers. See \cite{bazzi2016generative} for additional details about these benchmarks.

In Fig.~\ref{fig:benchmarkTemporal}, we show the performance of our iterative modularity-maximization algorithm on the uniform temporal benchmark from \cite{bazzi2016generative}. We use the same parameter values to generate networks that Bazzi et al. \cite{bazzi2016generative} used for their calculations. The iteration starts from $\gamma^{(0)}=1$ and $\omega^{(0)}=1$. The two plots use {\sc GenLouvain} and {\sc GenLouvainRand} to maximize modularity. {\sc GenLouvain} tends to perform better, especially for larger values of the copying probability $p$. The plots from Fig.~\ref{fig:benchmarkTemporal} have the same qualitative features as those from Fig.~\ref{fig:dynamicModel}, with NMI scores close to $1$ for $\mu$ close to $0$ and rapid degradation in performance for critical values of $\mu$ that depend on the value of the copying probability $p_b$. We also observe a clear separation of the curves that correspond to different values of $p$, with 
better performance as $p$ increases. In particular, when $p_b=1$, we see some level of recovery of the planted community structure up
to $\mu=1$. Towards the bottom of both plots, we show for each value of $p_b$ the value of $\mu$ up to which at least $10\%$ of the $100$ runs converge to a fixed point in a maximum of $30$ iterations. Recall that when our algorithm does not converge, it returns the largest-modularity partition that it encounters during the iterative process. For smaller values of $p_b$, the mean NMI between the algorithmically-detected and planted communities remains large beyond these critical values of $\mu$, suggesting the utility of our approach even when the iteration fails to converge. For $p_b=1$, the critical value of $\mu$ appears to coincide with the value of $\mu$ beyond which the calculated NMI is close to $0$. 

In \cite{bazzi2016generative}, Bazzi et al. fixed $\gamma=1$ and computed NMI between algorithmically-detected and planted communities for different values of $\omega$.
They did not attempt to select an ``optimal" set of parameter values, as this was not a goal of their paper. Using Alg.~\ref{alg:itModMax} to select values of $\gamma$ and $\omega$ in a principled way, we obtain NMI values that are comparable with or larger than those that were observed for the analogous calculations in \cite{bazzi2016generative} using the best possible value of $\omega$ (i.e., the one that yields the largest NMI between detected and planted partitions) \footnote{Bazzi et al. used a different version of NMI (one that normalizes the mutual information by the joint entropy of the input partitions). Thus, it is not possible to systematically compare our results from Fig.~\ref{fig:benchmarkTemporal} with theirs. However, we have recalculated some NMI values for specific choices of $p_b$ and $\mu$ to enable a direct comparison to \cite{bazzi2016generative}. To give an example, for $p_b=0.85$ and $\mu=0.6$, we obtain \unexpanded{$\langle \mbox{NMI} \rangle \approx 0.70$} using {\sc GenLouvain} and \unexpanded{$\langle \mbox{NMI} \rangle \approx 0.65$} using {\sc GenLouvainRand}. By comparison, Bazzi et al. obtained \unexpanded{$\langle \mbox{NMI} \rangle \approx 0.55$} and \unexpanded{$\langle \mbox{NMI} \rangle \approx 0.40$} for the two versions of the algorithm.}.  


\begin{figure}[hbt!]
	\centering
	\subfloat[First iteration]{\includegraphics[width=0.35\textwidth]{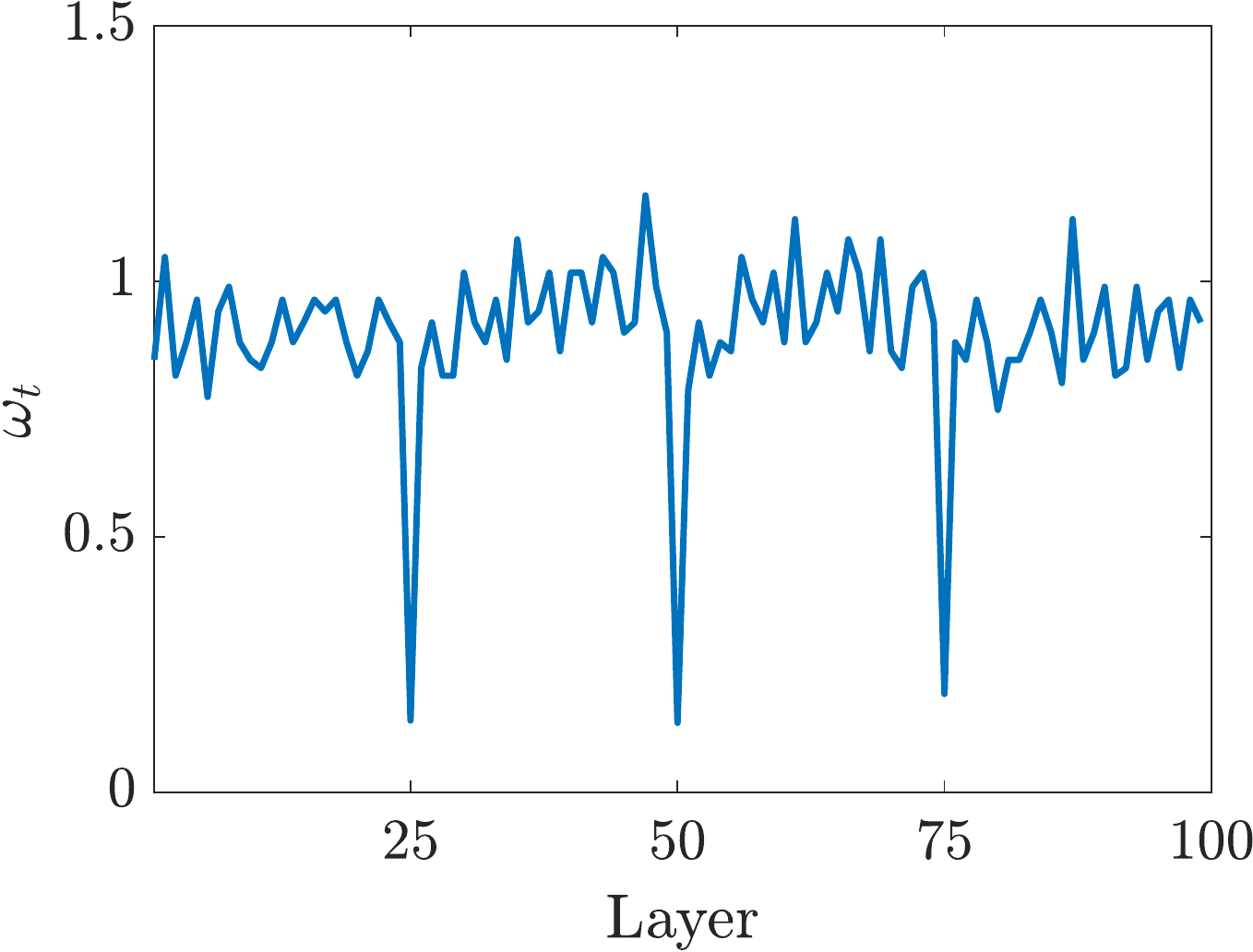}} \hspace{0.1\textwidth}  
	\subfloat[Final iteration]{\includegraphics[width=0.35\textwidth]{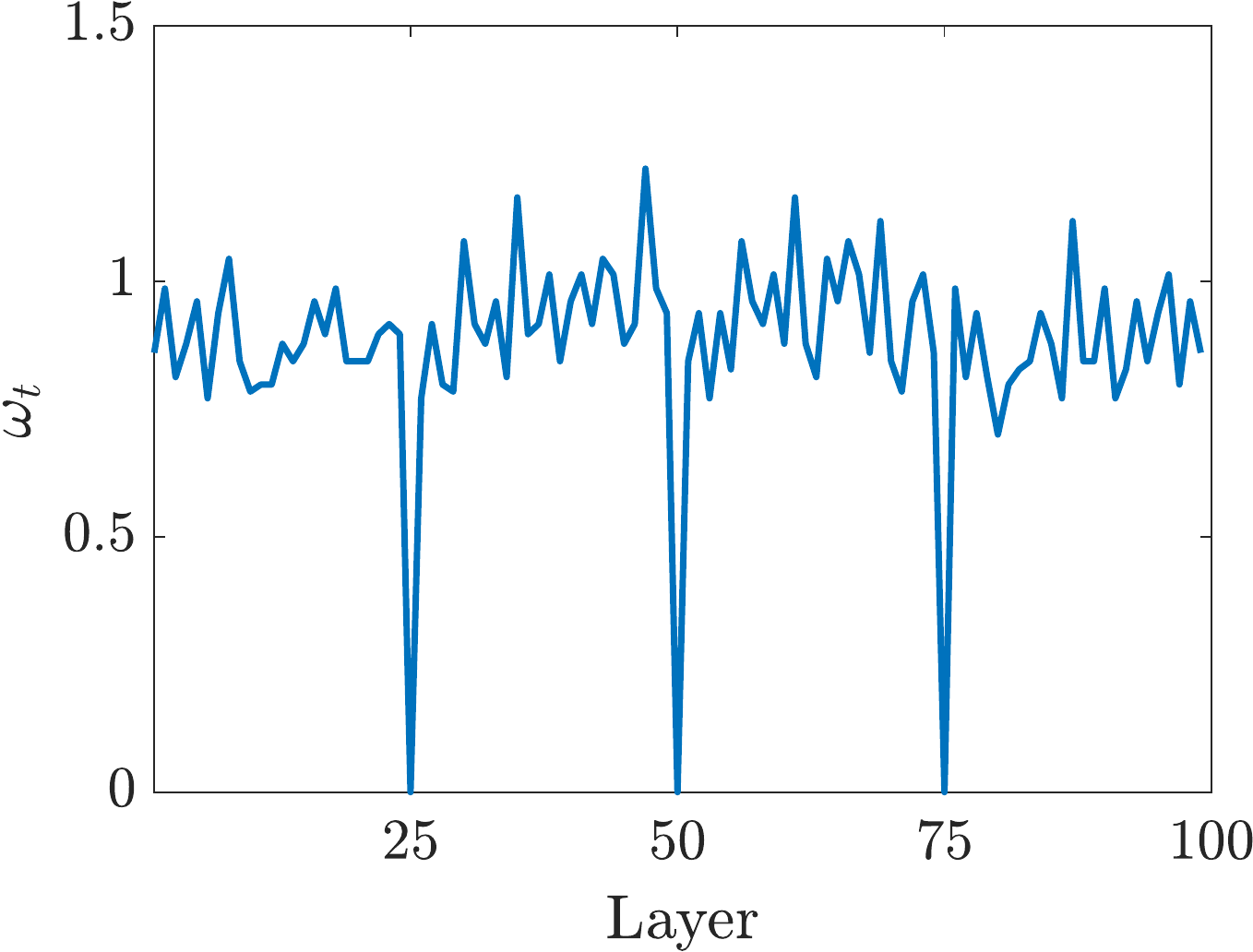}}
	\caption{Inferred values of the coupling parameters $\omega_t$ after (a) one iteration of our {\sc IterModMax} algorithm and (b) the final iteration (i.e., upon convergence) for a temporal network with change points. We use $p_t=0$ for $t \in \{25,50,75\}$; this induces an abrupt change in community structure at these layers. For the remaining layers, we set $p_t=0.9$. We let $\mu=0.4$ for the mixing parameter; other choices give similar results. For the remaining benchmark parameters, we use the same values as in \cite{bazzi2016generative}. Specifically, we generate a network with $T=100$ layers and $N=150$ nodes in each layer. There are $K=5$ communities, whose expected sizes follow a symmetric Dirichlet distribution with a concentration parameter with value $1$;
		and node degrees follow a truncated power-law distribution. (See \cite{bazzi2016generative} for details.)}
	\label{fig:benchmarkChangePoints}
\end{figure}

Bazzi et al. also examined temporal networks with non-uniform interlayer dependencies, with a focus on networks with change points \cite{bazzi2016generative}. We now apply our iterative algorithm with layer-dependent values of $\gamma$ and $\omega$ to one such network with $T=100$ layers using a mixing parameter $\mu=0.4$. We set the copying probability to be $p_t=0$ for $t \in \{25,50,75\}$ (the change-point layers), and we set $p_t=0.9$ for all other layers. This induces a reorganization of mesoscale structure in layers $25$, $50$, and $75$ of a sampled network. In Figure~\ref{fig:benchmarkChangePoints}, we show the inferred values of the coupling parameters $\omega_t$ using our layer-dependent {\sc IterModMax} algorithm both (a) after one iteration of the algorithm and (b) upon convergence (i.e., in the final iteration). Our algorithm correctly infers $\omega_t=0$ at the three change points. The layer-averaged NMI between the output partition and the planted partition is approximately $0.96$.

We make a few comments about Figure~\ref{fig:benchmarkChangePoints}. First, these results use a version of the {\sc IterModMax} algorithm that fixes $\gamma=1$ throughout the iterations. (Due to the large number of parameters in the layer-dependent case, Algorithm \ref{alg:itModMax} does not converge when we vary both $\gamma$ and $\omega$.)
We initialize all coupling parameters $\omega_t$ to $1$.
Second, we use a version of the {\sc GenLouvain} algorithm without the post-processing step that increases multilayer modularity by relabeling communities to increase persistence. (See \cite{bazzi2016} and the implementation of {\sc GenLouvain} in \cite{genlouvain} for details.) With post-processing, our algorithm is unable to determine that layers $24$ and $25$ have unrelated community structures (and likewise for the other two change points). Third, although it can take our {\sc IterModMax} algorithm up to $20$--$30$ iterations to converge to a solution like the one from Figure~\ref{fig:benchmarkChangePoints}(b), it only takes one iteration to identify the change-point layers [see Figure~\ref{fig:benchmarkChangePoints}(a)]. This suggests that waiting for the iterative algorithm to converge may not be necessary if one is interested only in the locations of change points (and not in the extent to which communities reorganize at change points).

\begin{figure}[htbp]
	\centering
	\subfloat[{\sc GenLouvain}]{\includegraphics[width=0.45\textwidth]{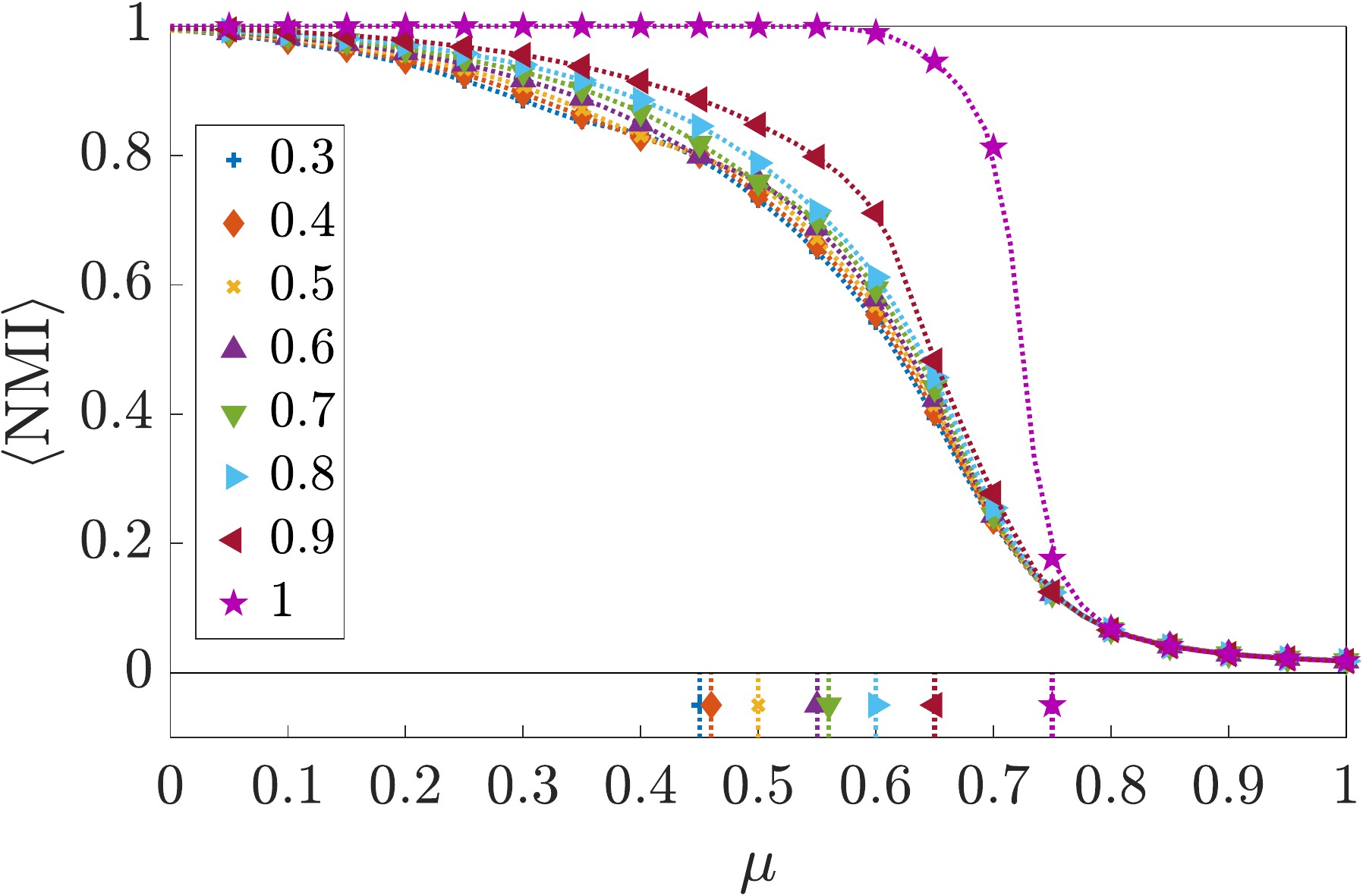}} \hfill
	\subfloat[{\sc GenLouvainRand}]{\includegraphics[width=0.45\textwidth]{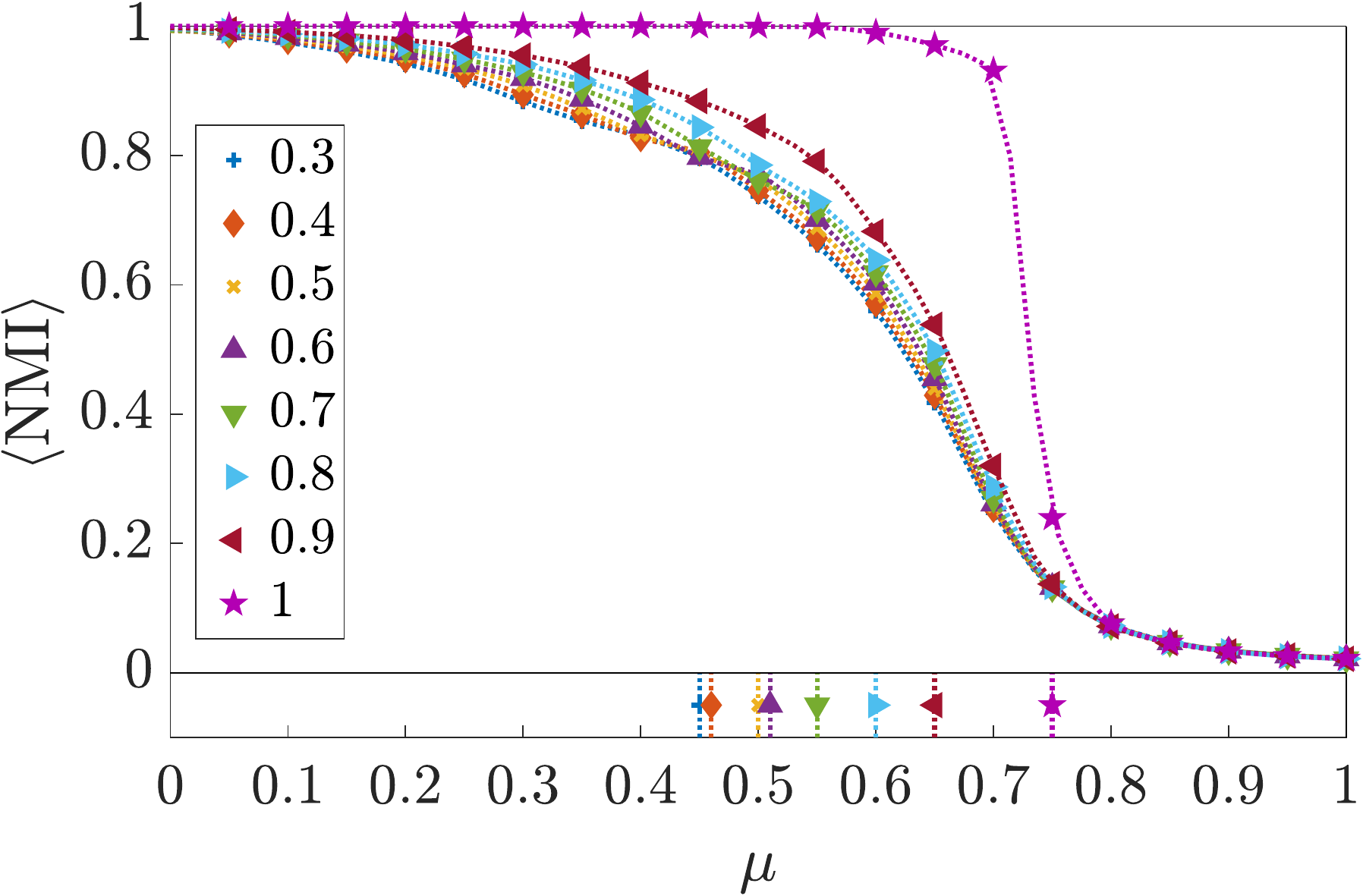}}
	\caption{Results for the multiplex benchmark from Bazzi et al. \cite{bazzi2016generative} using {\sc GenLouvain} and {\sc GenLouvainRand} to perform modularity maximization. We use the same parameter values that the authors used in their paper. The resulting networks have $T=15$ layers and $N=1000$ nodes in each layer. There are $K=10$ communities, whose expected sizes follow a symmetric Dirichlet distribution with 
		a concentration parameter with value $1$;
		and node degrees in each layer follow a truncated power-law distribution. (See \cite{bazzi2016generative} for details.) Our plots show layer-averaged NMI scores between the planted partition and the one that we detect by modularity maximization as a function of the mixing parameter $\mu$. Each line and set of markers corresponds to a different value of the copying probability $\widehat{p}_b=(T-1)p_b$, and each data point is a mean over $100$ trials. We use Alg.~\ref{alg:itModMax} to update the values of $\gamma$ and $\omega$ from the initial values
		$\gamma^{(0)}=1$ and $\omega^{(0)}=1/T$. The lines at the bottom of each plot indicate the values of $\mu$ up to which at least $10\%$ of the $100$ runs converge to a fixed point.}
	\label{fig:benchmarkMultiplex}
\end{figure}

In Fig.~\ref{fig:benchmarkMultiplex}, we show the performance of our method on the uniform multiplex benchmark network from \cite{bazzi2016generative}. As with the temporal benchmark, we use the same parameter values to generate networks that Bazzi et al. used in their numerical experiments. We initialize our iterative algorithm with $\gamma^{(0)}=1$ and $\omega^{(0)}=1/T$, where $T$ is the number of layers. (We justify our choice of $\omega^{(0)}$ using the scaling argument from Sec.~\ref{subsec:multiplex}.) For both {\sc GenLouvain} and {\sc GenLouvainRand}, and with the exception of the case with $\widehat{p}_b=1$, we observe only marginal improvement in performance for larger values of the copying probability $\widehat{p}_b$. Bazzi et al. made a similar observation in \cite{bazzi2016generative}. For $\widehat{p}_b=1$, the abrupt drop in performance coincides with the point at which fewer than $10\%$ of runs converge to a fixed point. (Our choice of threshold is arbitrary; the results are similar for values other than 10\%.) Our method of estimating the parameters $\gamma$ and $\omega$ gives comparable NMI values to the best values that Bazzi et al. obtained in \cite{bazzi2016generative} for their analogous calculations \footnote{To be consistent with \cite{bazzi2016generative}, we calculate a second set of NMI values, which normalize mutual information by the joint entropy of the input partitions, rather than by their mean entropy.}.

Our numerical experiments in this section demonstrate the promise of our approach for estimating resolution and interlayer-coupling parameters. We obtain NMI scores that often exceed the best NMI values that have been reported for multilayer modularity maximization without the parameter-estimation step \cite{bazzi2016generative}. It is inevitable that, for some networks, modularity maximization will underperform other community-detection methods \cite{peel2017ground} (e.g., the belief-propagation algorithm from \cite{ghasemian2016} or inference algorithms based on general SBMs). However, when using
modularity maximization 
for performing community detection, our way of estimating resolution and coupling parameters is a valuable tool.

\bibliography{ref}


\end{document}